\documentclass[aps,prb,reprint,amsmath,amssymb,superscriptaddress,nofootinbib]{revtex4-2}

\usepackage{graphicx}
\usepackage{dcolumn}
\usepackage{braket}
\usepackage{bm}
\usepackage[breaklinks=true,colorlinks,citecolor=blue,linkcolor=blue,urlcolor=blue]{hyperref}
\usepackage{enumitem}
\usepackage{subfigure}
\usepackage{threeparttable}
\begin{document}
\title{Neural Network Evolution Strategy for Solving Quantum Sign Structures}

\author{Ao Chen}
 \email{aochen@student.ethz.ch}
\affiliation{Department of Physics, ETH Zurich, CH-8093 Zurich, Switzerland}

\author{Kenny Choo}
\affiliation{Department of Physics, University of Zurich, CH-8057 Zurich, Switzerland}

\author{Nikita Astrakhantsev}
\affiliation{Department of Physics, University of Zurich, CH-8057 Zurich, Switzerland}

\author{Titus Neupert}
\affiliation{Department of Physics, University of Zurich, CH-8057 Zurich, Switzerland}

\begin{abstract}
Feed-forward neural networks are a novel class of variational wave functions for correlated many-body quantum systems. 
Here, we propose a specific neural network ansatz suitable for systems with real-valued wave functions. Its characteristic is to encode the all-important rugged sign structure of a quantum wave function in a convolutional neural network with \emph{discrete} output. Its training  is achieved through an evolutionary algorithm. 
We test our variational ansatz and training strategy on two spin-1/2 Heisenberg models, one on the two-dimensional square lattice and one on the three-dimensional pyrochlore lattice. In the former, our ansatz converges with high accuracy to the analytically known sign structures of ordered phases. In the latter, where such sign structures are a priory unknown, we obtain better variational energies than with other neural network states. 
Our results demonstrate the utility of discrete neural networks to solve quantum many-body problems.
\end{abstract}

\maketitle

\textit{Introduction.--\,} Finding the ground states of frustrated quantum magnets poses a formidable challenge to existing numerical algorithms. 
For instance, exact diagonalization (ED) methods are limited to small system sizes, while quantum Monte Carlo (QMC)~\cite{Ceperley_Science86_QMC} fails in frustrated models due to the notorious sign problem~\cite{Troyer_PRL05_SignProblem}. The density matrix renormalization group (DMRG)~\cite{Verstraete_AiP08_DMRG, Schollwock_AP11_MPS}, despite being very successful in low-dimensional models, struggles to express entanglement in higher-dimensional systems. 
Complementary to all these are variational approaches, which are in principle scalable to larger systems, but carry different variational bias~\cite{Becca_17_QMCtext}. To reduce the latter, many-variable schemes have been put forward, either based on pair product wave functions~\cite{Tahara_JPSJ08_mVMC} or on application of neural networks (NN) to quantum systems known as {\it Neural Quantum States} (NQS) approach~\cite{Carleo_Science17_NQS}.

Numerous studies aimed at improving the accuracy of the NQS wave functions in order to take full advantage of their expressibility for many-body calculations. For instance, incorporating point group symmetries of the state into the restricted Boltzmann machine (RBM), employing the convolutional neural network (CNN) \cite{Dumoulin_arxiv18_IntroCNN} or the group-CNN \cite{Cohen_arxiv16_GCNN} structure has led to significant improvements in the performance of NQS~\cite{Nomura_JPCM2021_RBMsymm, Choo_PRB19_J1J2CNN, Roth_arxiv21_GCNN}. The $SU(2)$ symmetry can also be enforced to improve the approximation quality~\cite{Vieijra_PRL20_NQSsymm, Vieijra_arxiv21_NQSsymm}. The large-scale implementation of deep NNs is another way to obtain better variational states~\cite{Sharir_PRL20_QVAN, Li_PRR20_ISGO, Li_arxiv21_LargeScaleCNN}.
Furthermore, combining several NNs~\cite{Cai_PRB18_CombinedNet, Szabo_PRR20_SignProblem}, NNs with traditional Gutzwiller-projected many-variable wave functions~\cite{Ferrari_PRB19_GWFRBM, Nomura_PRB17_RBMPP, Nomura_arxiv20_J1J2RBMPP} and NNs with projected entangled pair states~\cite{Liang_PRB21_CNNPEPS} has led to more accurate variational sign structures and better variational solutions.

In spite of all these improvements, representing and finding sign structures is still a major problem in pure NQS methods~\cite{Westerhout_NC20_FrustratedDifficulty}, which prevents NQS from obtaining desirable variational states in highly frustrated systems. For instance, in three-dimensional systems such as the frustrated Heisenberg pyrochlore model, NQS methods obtained less accurate variational energies than the pair product wave function ansatz~\cite{Astrakhantsev_arxiv21_pyrochlore}. 

To alleviate this problem of sign structure approximation, we propose a new architecture constructed by a combination of a base network and a sign `helper' network. The sign network has discrete output values representing signs of wave function amplitudes. Such discrete sign network requires a gradient-free optimization method~\cite{Audet_17_blackbox}. In this work, we adopt an evolutionary algorithm -- the evolution strategy (ES)~\cite{Hansen_arxiv16_CMAES} -- to perform the optimization. We contrast our ansatz with a previously introduced one, where a helper network that represents a continuous complex phase was introduced~\cite{Szabo_PRR20_SignProblem}. 

To demonstrate an improvement over existing NQS techniques, we test the performance of our network with ES optimization on the spin-1/2 $j_1$--$j_2$ Heisenberg models on both the two-dimensional (2D) square lattice and the three-dimensional (3D) pyrochlore lattice. Both of these are exemplary open problems in frustrated quantum magnetism. In the former system, we illustrate the working mechanism of the combined network and the advantage of introducing the ES optimization. In the latter, we show that our approach can produce competitive variational states with variational energies that are better than all previous NQS studies~\cite{Astrakhantsev_arxiv21_pyrochlore} and are comparable to the ones obtained within DMRG~\cite{Hagymasi_PRL21_PyrochloreDMRG}. Finally, we discuss the optimal NN architecture for learning wave function sign structures and the future prospects of the ES approach.

\textit{Sign network with evolution strategy.--\,} We use a product of two separate networks to express the variational wave function $\psi$ as 
\begin{equation} \label{eq:wave_function}
    \psi(\sigma) = \psi_{\mathrm{b}}(\sigma) \times \psi_{\mathrm{h}}(\sigma),
\end{equation}
where $\sigma$ is the spin configuration of a spin-1/2 magnet in a fixed basis, while $\psi_{\mathrm{b}}$ and $\psi_{\mathrm{h}}$ are the outputs of the base and helper networks, respectively. 
The details of the base network are given in Appendix~\ref{base network}.

\begin{figure}[t]
    \centering
    \includegraphics[width=0.45\textwidth]{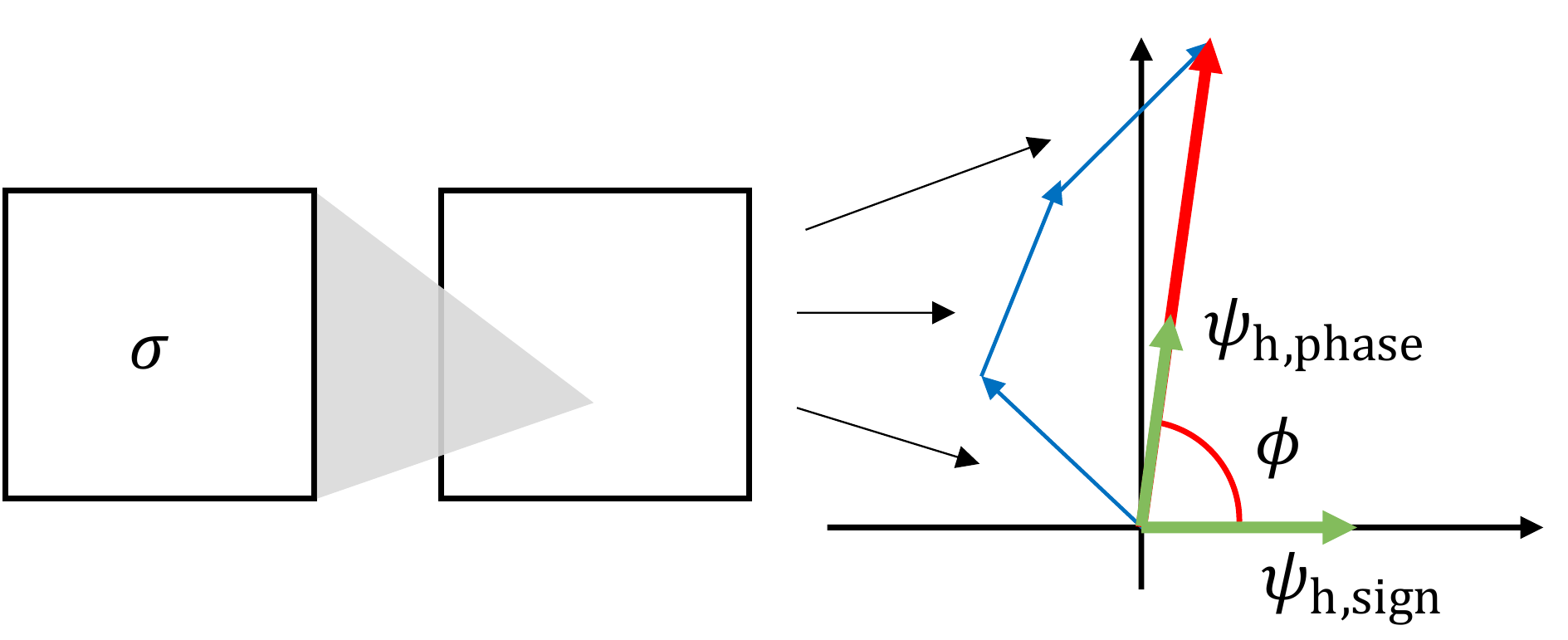}
    \caption{Architecture of the helper network. The network has a convolutional layer followed by an activation function which returns a phase $\phi$. One convolutional layer only contains one channel without bias. The red vector represents the sum of all blue vectors and the green vectors are the outputs of phase and sign networks.}
    \label{fig:helper_network}
\end{figure}

The purpose of introducing an additional helper network is to help the base network to express the sign structure and increase the accuracy of the variational state. The helper network has a one-layer convolutional architecture, as shown in Fig.\,\ref{fig:helper_network}. To interpret the output of the convolutional layer as sign structure, we apply the approach used in Ref.\,\cite{Szabo_PRR20_SignProblem}. Namely, every element in the output of the slngle convolutional layer is treated as the phase of a unit complex number (phasar), and the sum of all these phasars gives the output phase, i.\,e.
\begin{equation}
\label{eq:phase}
    \phi(\sigma) = {\mathrm{arg}} \left( \sum_j e^{i v_j(\sigma)} \right),
\end{equation}
where $v_j$ is the output of the convolutional layer. In the previous NQS studies, the helper network usually expresses the phase of wave functions $\psi_{\mathrm{h, phase}}(\sigma) = e^{i\phi(\sigma)}$ which yields a continuous output~\cite{Szabo_PRR20_SignProblem, Torlai_NP18_Tomography, Astrakhantsev_arxiv21_pyrochlore}. Here, we are interested in systems for which the wave functions can be chosen to be real-valued. For these cases $\psi_{\mathrm{h, phase}}(\sigma)$ introduces an unnecessary redundancy which can potentially reduce the performance. In contrast, the output of the sign helper network is given by
\begin{equation}
\label{eq:sign_network}
    \psi_{\mathrm{h, sign}} (\sigma) = {\mathrm{sgn}}\left[\cos\phi(\sigma)\right],
\end{equation}
which maps the phase $\phi$ to a discrete sign $\pm 1$ and eliminates the redundancy. 

\begin{figure}[t]
    \centering
    \subfigure[Continuous case]{
        \includegraphics[width=0.45\textwidth]{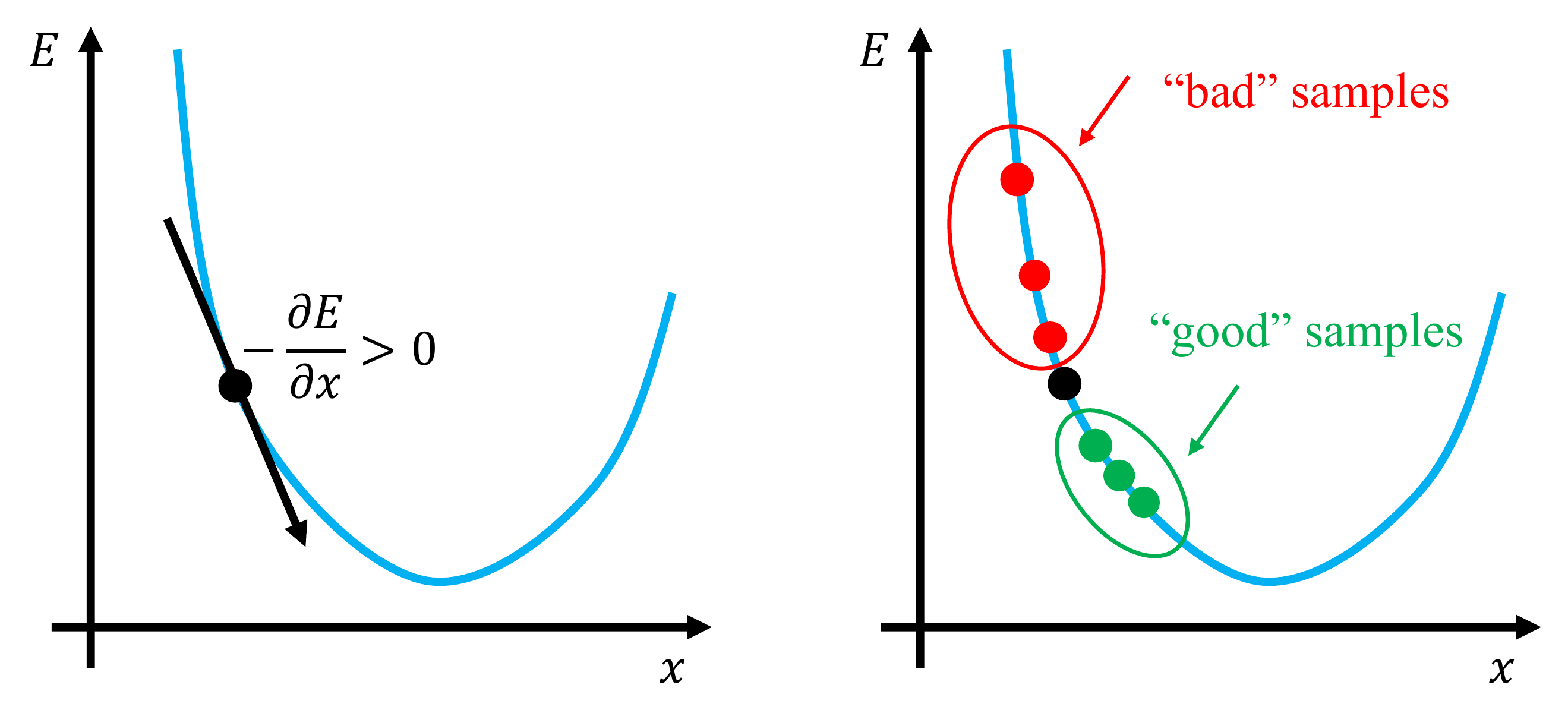}
    }
    \hfill
    \subfigure[Discrete case]{
        \includegraphics[width=0.45\textwidth]{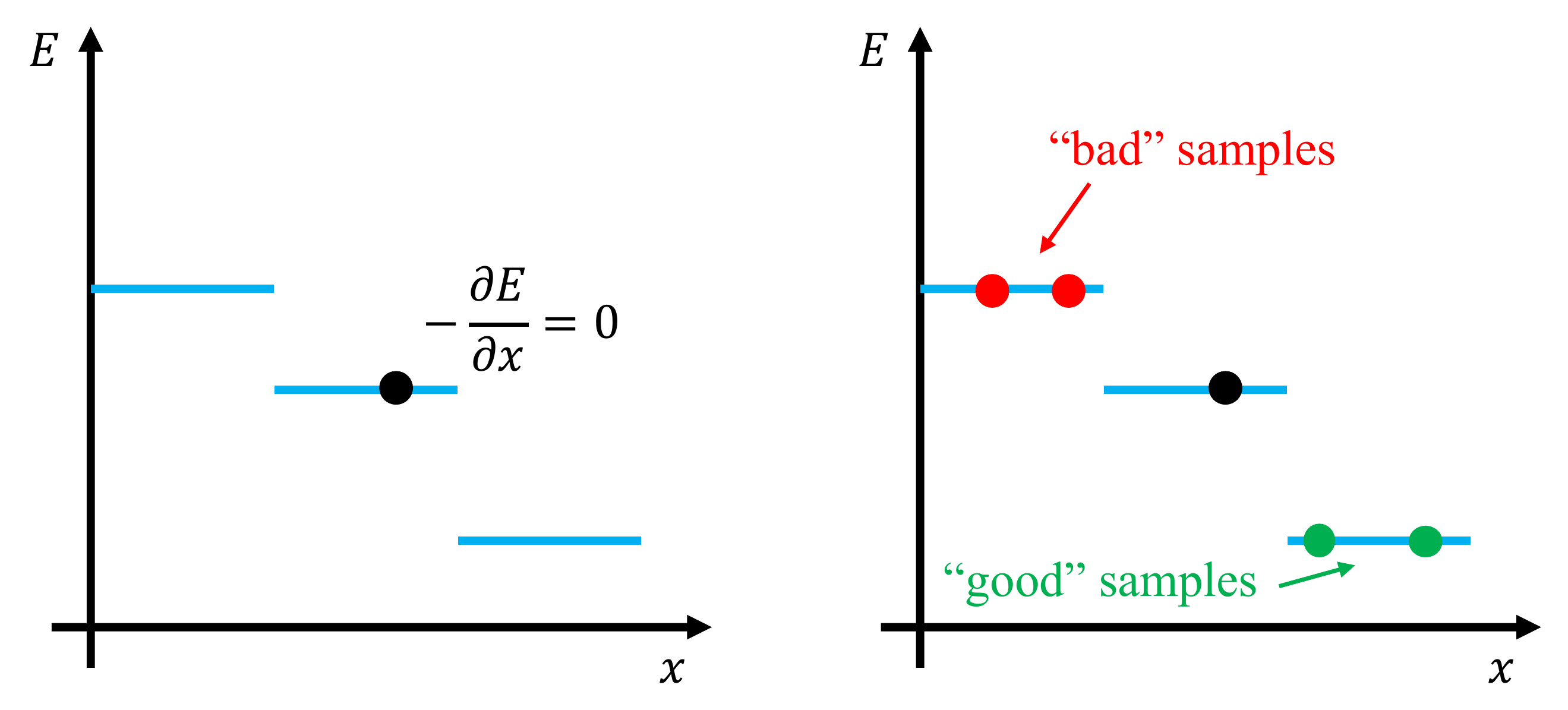}
    }
    \caption{Comparison between gradient descent and evolution strategy (ES). Both of them are applicable to minimize continuous functions, but only ES is applicable in the discrete case.}
    \label{fig:sgd_es}
\end{figure}

We adopt the gradient-free ES method to train the sign network, while the base network and the phase helper network (required for performance comparison) are trained using stochastic reconfiguration (SR)~\cite{Sorella_JCP07_SR}. Compared with other gradient-free methods, the ES has shown outstanding performance in optimizing complicated functions~\cite{hansen_10_compare}. In Fig.\,\ref{fig:sgd_es} we illustrate the difference between the ES and gradient-based optimization methods such as stochastic gradient descent (SGD). If the energy $E$ is a continuous function of a variable $x$, SGD moves $x$ in the steepest descent direction $-\frac{\partial E}{\partial x}$. Instead, the ES method generates samples in the vicinity of the current variable value $x$ and estimates the direction of energy descent therewith. When optimizing continuous functions, the ES method has no advantage as compared to SGD. However, if the function is piecewise constant in the variational parameters, which is the case with the sign network introduced in Eq.\,\eqref{eq:sign_network}, the ES can be used to optimize the variational wave function, while SGD is no longer applicable.

There exist numerous variations of the ES optimization method. In this paper, we employ the covariance matrix adaptation evolution strategy (CMA-ES)~\cite{Hansen_arxiv16_CMAES}. This iterative algorithm consists of the following three steps in each iteration:
\begin{enumerate}[itemsep=0pt, topsep=0pt]
    \item Starting from the sign network weights $\bm W$, generate a set of sample networks with different weights drawn from the multivariate normal distribution 
    \begin{equation}
        \bm X_n \sim \mathcal{N} (\bm W, \Sigma) \sim \bm{W} + \sqrt{\Sigma} \, \mathcal N(0, \bm{1}),
    \end{equation}
    where the index $n$ labels sample networks and $\Sigma = \braket{(\bm X - \bm W)(\bm X - \bm W)^{\mathrm{T}}}$ is the covariance matrix.
    \item Compute the variational energy $E_n$ of these sample networks using reweighting Monte Carlo in Appendix \ref{reweighting}.
    \item Update the sign network weights as 
    \begin{equation} \label{eq:W_update}
        \bm W' = \bm W + \sum_{n=1}^N w_n (\bm X_n - \bm W)
    \end{equation}
    and the covariance matrix as 
    \begin{equation} \label{eq:Sigma_update}
        \Sigma' = \sum_{n=1}^N w_n (\bm X_n - \bm W)(\bm X_n - \bm W)^{\mathrm{T}},
    \end{equation}
    where $N$ is the number of samples. The samples with lower energy $E_n$ are assigned a higher importance $w_n$ (see Appendix \ref{CMA-ES} for details).
\end{enumerate}

\begin{figure}[t]
    \centering
    \includegraphics[width=0.45\textwidth]{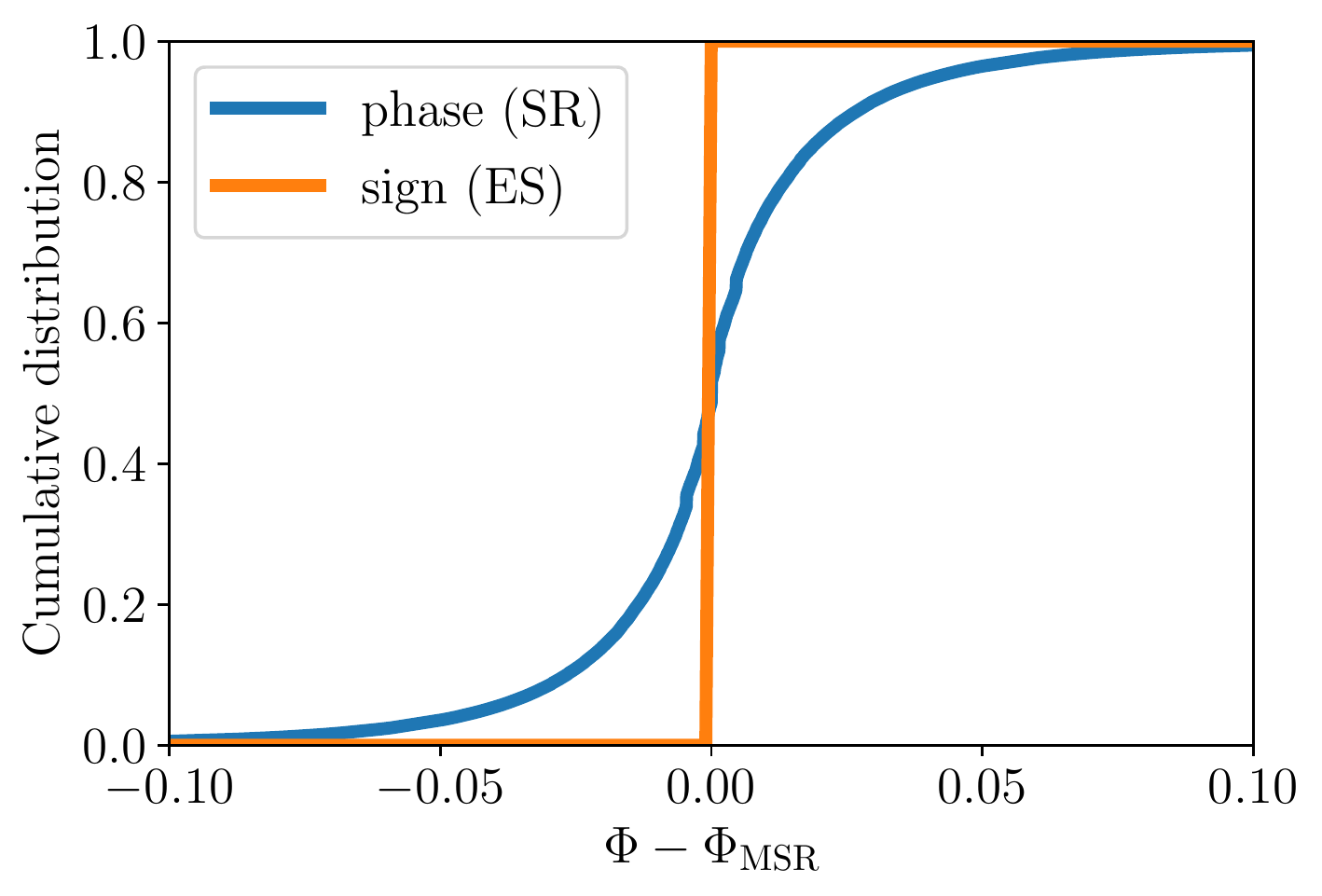}
    \caption{Cumulative distribution function (CDF) of phase relative to the MSR on the $10\times10$ square lattice. The sign network learns all signs correctly in $10^6$ samples, resulting into the step-like CDF with the step at $\Phi - \Phi_{\mathrm{MSR}} = 0$.}
    \label{fig:square_CDF}
\end{figure}

These iterations are repeated until the variational energy is converged. The \texttt{pycma} package~\cite{Hansen_19_pycma} is adopted to perform the CMA-ES optimization. Further details of CMA-ES are given in Appendix~\ref{CMA-ES}.

\textit{Numerical experiments.--\,} As benchmark examples, we study the spin-1/2 $j_1$--$j_2$ Heisenberg model on the 2D square lattice and on the 3D pyrochlore lattice. Both of these systems are possible hosts to quantum spin liquid states and have recently been subject to many-variable Monte Carlo and NQS variational studies, which led to a better understanding of their phase diagram~\cite{Nomura_arxiv20_J1J2RBMPP, Astrakhantsev_arxiv21_pyrochlore}. The Heisenberg Hamiltonian is given by
\begin{equation}
    H = j_1 \sum_{\braket{i,j}} {\bf{\hat S}}_i \cdot {\bf{\hat S}}_j 
    + j_2 \sum_{\left<\left<i,j\right>\right>} {\bf{\hat S}}_i \cdot {\bf{\hat S}}_j,
\end{equation}
where $\braket{...}$ and $\left<\left<...\right>\right>$ indicate nearest and next-to-nearest neighbour spins on the respective lattice. The maximally frustrated region is located in the vicinity of $j_2 / j_1 = 0.5$ for the square lattice~\cite{Gong_PRL14_J1J2DMRG, Nomura_arxiv20_J1J2RBMPP} and at $j_2 / j_1 = 0$ for the pyrochlore lattice~\cite{Iqbal_PRX19_Pyrochlore, Astrakhantsev_arxiv21_pyrochlore, Hagymasi_PRL21_PyrochloreDMRG}.

\begin{figure}[t]
    \centering
    \includegraphics[width=0.48\textwidth]{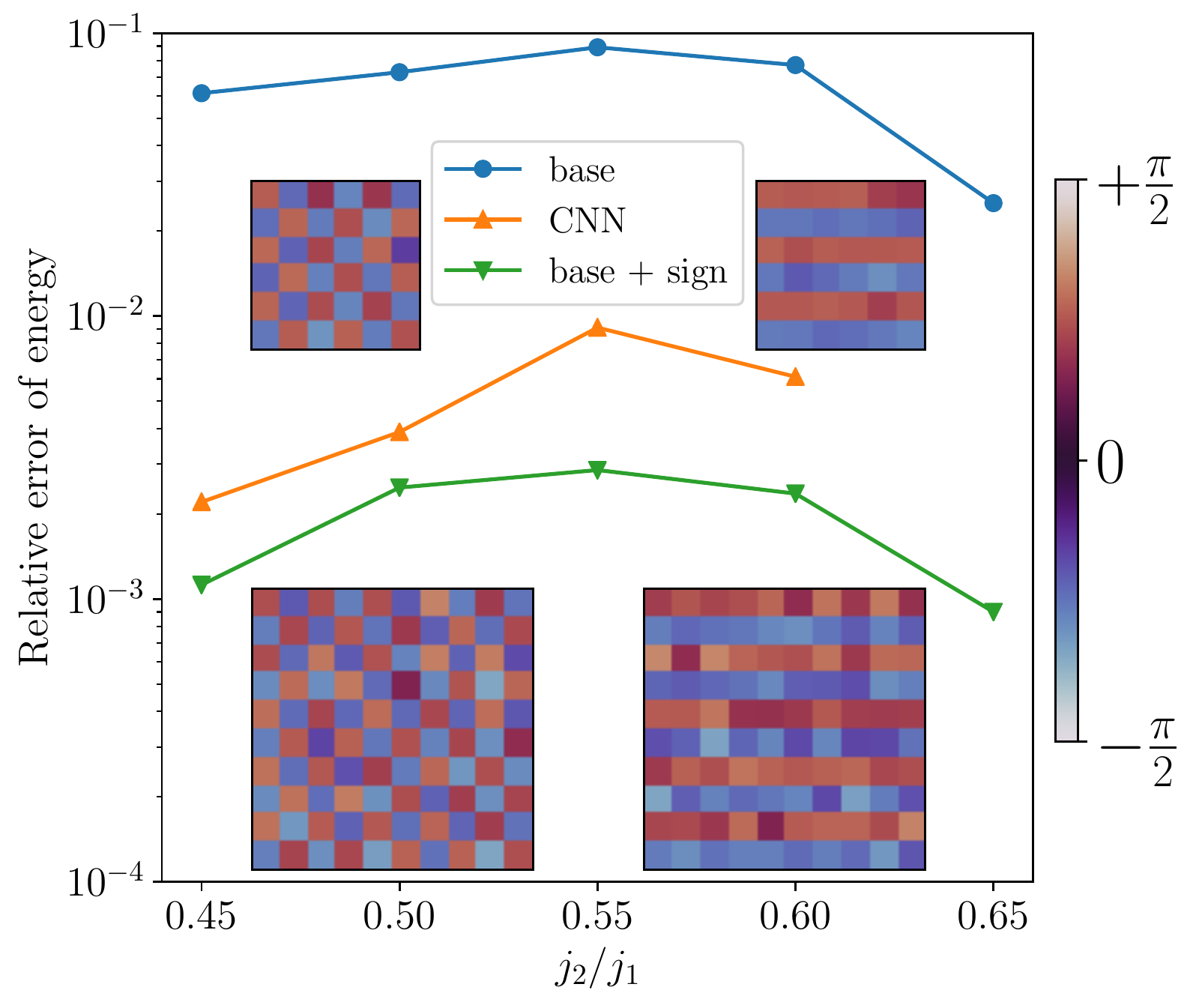}
    \caption{Relative error of the variational energy for the spin-1/2 $j_1$--$j_2$ Heisenberg model on the $6\times6$ square lattice. The ``base'' and ``base + sign'' points represent the performance of the base network without and with the sign network, respectively. The ``CNN'' data represents the approach employed in Ref.\,\cite{Choo_PRB19_J1J2CNN} which uses a complex-valued CNN to express the amplitude and phase with the MSR applied manually. The energy on larger lattices  (e.g., $10\times10$) is not shown because the base network cannot approximate the ground state without the sign network in this case. The insets show the weights of the sign network at $j_2/j_1 = 0.5$ and $j_2/j_1 = 0.6$ on the $6\times6$ and $10\times10$ lattices.} 
    \label{fig:pattern}
\end{figure}

In the case of even lattice dimensions in both directions, the square lattice can be divided into A and B sublattices with a checkerboard pattern or a stripe pattern. In the $j_2 / j_1 = 0$ and $j_1 / j_2 = 0$ limits, the sign of a wave function of a fixed spin configuration is given by $(-1)^{M_{\mathrm{A}}}$ with $M_{\mathrm{A}}$ being the total number of spin-up in the A sublattice for this configuration. In the $j_2 / j_1=0$ limit, the sublattice is a checkerboard pattern formed by one of the two bipartite components of the lattice. This special case is known as the Marshall sign rule (MSR)~\cite{Marshall_PRSA55_MSR}. In the $j_1 / j_2=0$ limit, a sublattice has a stripe pattern consisting of alternating rows or columns.

We start to test our approach in a case where the exact sign structure is known, the $10 \times 10$ square lattice at $j_2 / j_1 = 0$. The MSR in this limit, despite being easy to express analytically, can be tough for a generic NQS to learn and express~\cite{Liang_PRB18_CNNJ1J2}. To show how our method approximates the MSR, we subtract the phase $\Phi$ of the helper network outputs $\psi_\mathrm{h}(\sigma)$ from the phase $\Phi_{\mathrm{MSR}}$ given by the MSR.
In Fig.\,\ref{fig:square_CDF} we show the cumulative distribution function (CDF) of this difference in phases. The sign network trained by the ES leads to a step-function CDF, indicating that our approach finds the exact sign structure. We contrast this with the performance of the phase helper network in Ref.\,\cite{Szabo_PRR20_SignProblem} which expresses a continuous phase instead of a discrete sign. As seen from Fig.\,\ref{fig:square_CDF}, this network also approximates MSR, but contains significant phase errors\footnote{To reduce the phase error, a multi-channel phase network much more complex than the 1-channel helper network in Fig.\,\ref{fig:helper_network} is adopted in Ref.\,\cite{Szabo_PRR20_SignProblem}.}. This showcases that the discrete sign network trained by ES obtains a more accurate sign structure than the conventional continuous-phase networks for comparable network complexity.

We now consider the performance of the sign network in the frustrated region of the square lattice $j_1$--$j_2$ model. To this end, we show the relative error of energy and depict the sign network weights for various $0.45 \leqslant j_2/j_1 \leqslant 0.65$ in Fig.\,\ref{fig:pattern}. These weights explicitly show patterns corresponding to the checkerboard or the stripe patterns described above and represent the simple sign structures at the two extreme bipartite limits of Néel and stripe orders, which means the sign network can adapt to suitable simple sign structures successfully. Nevertheless, the sign network cannot express the remaining frustrated sign structure. This observation extends to more complex sign network architectures with, for instance, multiple channels and was also observed in Ref.~\cite{Szabo_PRR20_SignProblem}.
Although the sign network cannot express the full sign structure, the simple sign structures provided by the sign network still lead to significantly improved performance of the base network compared to traditional CNN approach in Ref.\,\cite{Choo_PRB19_J1J2CNN} (see Fig.\,\ref{fig:pattern}). The similar conclusion also holds for the $10\times10$ lattice, in which case the base network cannot even be trained to approximate the ground state without the sign network.

\begin{table}[t]
\centering
\caption{Overlap with the exact ground state and relative error of energy on the $4 \times 2^3$ pyrochlore Heisenberg model.}
\label{tab:pyrochlore_comparison}
    \begin{threeparttable}
    \begin{tabular}{c c c c c} 
        \hline
        Sign\tnote{a} & Amplitude\tnote{a}
        & Helper network & $|\mathrm{Overlap}|^2$ & $|\Delta E/E_\mathrm{GS}|$ \\
        \hline
        Variational & Exact & No & 0.87 & --- \\
        Variational & Exact & Sign network & 0.91 & ---\\
        Exact & Variational & No & 0.79& --- \\ 
        Exact & Variational & Sign network & 0.83 & ---\\
        Variational & Variational & No & 0.77 & $6.2\times10^{-3}$ \\
        Variational & Variational & Sign network & 0.82 & $4.1\times10^{-3}$ \\
        \hline
    \end{tabular}
    \begin{tablenotes}
        \item [a] The exact sign or exact amplitude represent the variational wave function expressed by the network with signs or amplitudes replaced by the exact ground state values.
    \end{tablenotes}
    \end{threeparttable}
\end{table}

The application of the sign network in the 2D square $j_1$--$j_2$ model might be considered a special case since the well-known MSR can be used as a good starting point~\cite{Choo_PRB19_J1J2CNN, Nomura_JPCM2021_RBMsymm} instead of using a sign network trained by ES. To complement this, we study the case of the 3D pyrochlore lattice in which such prior sign structures are unknown.

We begin with the performance assessment for our approach on the $4\times2^3$ ED-accessible pyrochlore lattice cluster. In Table\,\ref{tab:pyrochlore_comparison}, we present the relative error of variational energy and the squared overlap between the variational state and the exact ground state to compare the improvement provided by the sign helper network.
The data with variational amplitudes and exact signs represents the variational state with the sign structure replaced by the exact ground state sign structure, and hence reflects the accuracy of the variational amplitude. Similarly, the  data with variational signs and exact amplitudes reflects the accuracy of the variational sign structure. These results show that the sign helper network not only permits more accurate sign structures, but also helps the base network to learn more accurate amplitudes. The improvement in the signs and amplitudes altogether leads to better wave functions as shown by the energy error and the overlap of the variational wave function. The supplementary data for the performance of different approaches on the $4\times2^3$ pyrochlore lattice is given in Appendix \ref{supplementary_pyrochlore}.

\begin{figure}[t]
    \centering
    \includegraphics[width=0.48\textwidth]{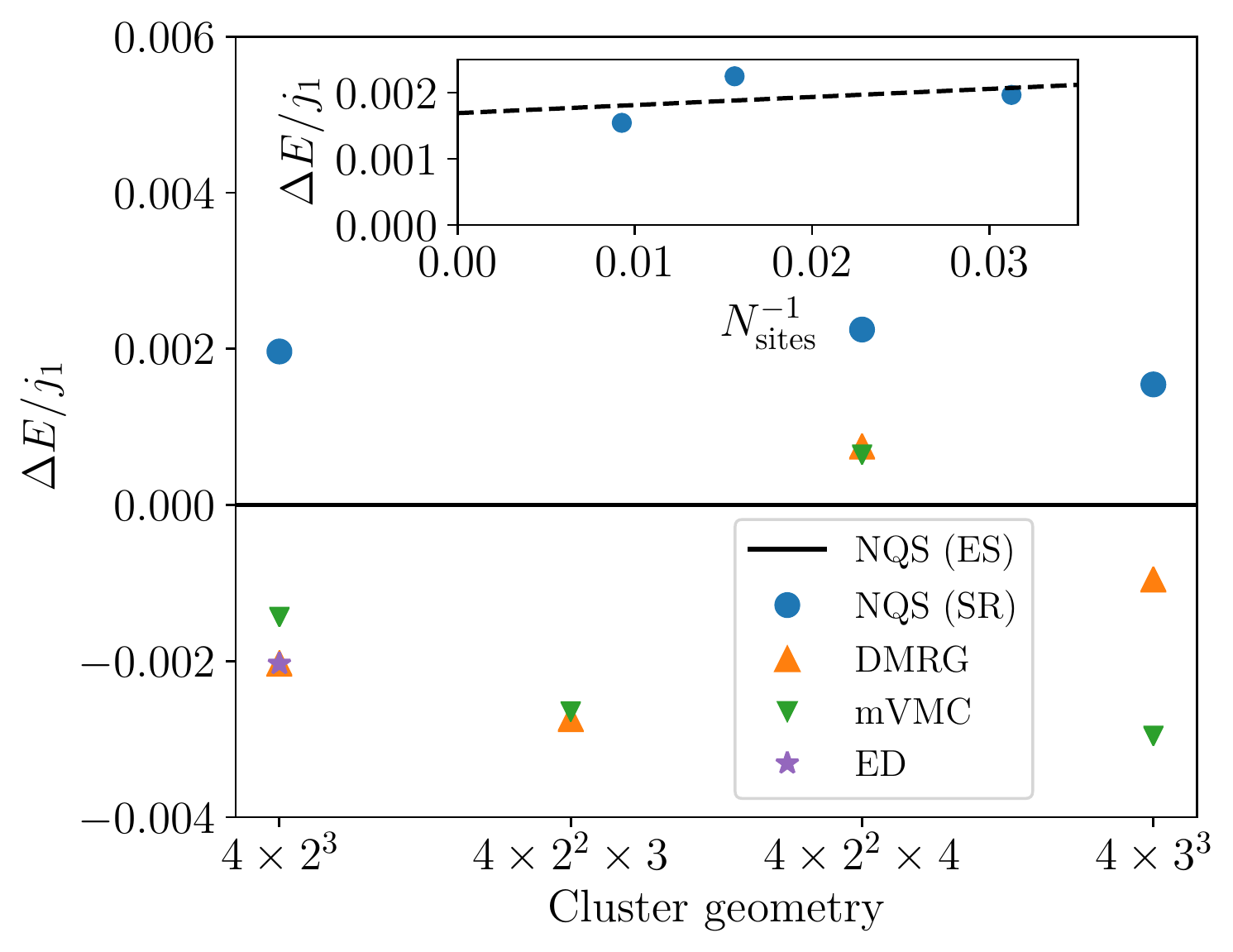}
    \caption{Energy difference with respect to the ES approach in the pyrochlore Heisenberg model at various lattice sizes. The inset shows the infinite-size extrapolation for the NQS (SR) data. The NQS (SR) approach in Ref.\,\cite{Astrakhantsev_arxiv21_pyrochlore} contains two separate CNNs for amplitude and phase. The mVMC and ED data are also given in Ref.\,\cite{Astrakhantsev_arxiv21_pyrochlore}. The DMRG data is given in Ref.\,\cite{Hagymasi_PRL21_PyrochloreDMRG}.}
    \label{fig:energy_difference}
\end{figure}

In Fig.\,\ref{fig:energy_difference}, we compare our approach with other numerical methods in the pyrochlore Heisenberg model at various lattice sizes\footnote{The same symmetry group projector and similar amount of parameters compared with Ref.\,\cite{Astrakhantsev_arxiv21_pyrochlore} is used to ensure a fair comparison between the two approaches.}. The performance of the base network without a sign helper network is not shown because its variational energy is far higher than other data exhibited in the plot when the lattice size is larger than $4\times2^3$. The base network with the sign helper network trained by ES improves the result of the pure SR method on all available cluster sizes with a non-vanishing energy gain under infinite-size extrapolation. This shows that the ES approach is in principle scalable to larger system sizes. Furthermore, our method also approaches the accuracy of DMRG for sizes larger than $4\times2^2\times3$. These numerical results show that the combination of the base and sign networks can improve the performance of NQS to a level comparable to other modern numerical methods in highly frustrated systems and scale the approach to larger volumes.

\textit{Discussion.--\,} In this work, we introduced a novel NQS ansatz for quantum many-body spin systems that consists of a continuous and discrete part, the latter of which is trained with an evolutionary algorithm. The intrinsic discrete nature of quantum sign structures provides an ideal application scenario for such a derivative-free optimization method. The ES allows us to use the non-differentiable sign network as a part of variational ansatz, which is a new numerical approach for quantum many-body problems. The advantage brought by ES is the elimination of all redundant phase factors that lead to a poorer optimization. The success of the non-differentiable sign network obtained in this paper hints at a possible prospect of other non-differentiable variational states, which may give more insights in the study of frustrated systems.

Furthermore, we found that a reliable network architecture to learn the sign structure is the combination of a base network and a sign helper network. 
The sign network, with an independent sign structure uninfluenced by the amplitude, is successful at expressing basic sign structures such as MSR. On the other hand, the base network also expresses a part of the sign structure together with the amplitude, which is indispensable in some frustrated systems as shown in our numerical experiments. Consequently, future NQS studies could focus on the combined network consisting of a base network and a sign helper network, and the ES method proposed in this paper provides a reliable approach for optimizing the primary sign structure expressed by the sign network in various situations.

\textit{Acknowledgments.--\,} T.\,N., K.\,C., and N.\,A. acknowledge support from the Swiss National Science Foundation (grant number: PP00P2$\_$176877) and from the European Research Council (ERC) under the European Union’s Horizon 2020 research and innovation program (ERC-StG-Neupert-757867-PARATOP). 

\bibliography{reference}

\begin{thebibliography}{42}%
\makeatletter
\providecommand \@ifxundefined [1]{%
 \@ifx{#1\undefined}
}%
\providecommand \@ifnum [1]{%
 \ifnum #1\expandafter \@firstoftwo
 \else \expandafter \@secondoftwo
 \fi
}%
\providecommand \@ifx [1]{%
 \ifx #1\expandafter \@firstoftwo
 \else \expandafter \@secondoftwo
 \fi
}%
\providecommand \natexlab [1]{#1}%
\providecommand \enquote  [1]{``#1''}%
\providecommand \bibnamefont  [1]{#1}%
\providecommand \bibfnamefont [1]{#1}%
\providecommand \citenamefont [1]{#1}%
\providecommand \href@noop [0]{\@secondoftwo}%
\providecommand \href [0]{\begingroup \@sanitize@url \@href}%
\providecommand \@href[1]{\@@startlink{#1}\@@href}%
\providecommand \@@href[1]{\endgroup#1\@@endlink}%
\providecommand \@sanitize@url [0]{\catcode `\\12\catcode `\$12\catcode
  `\&12\catcode `\#12\catcode `\^12\catcode `\_12\catcode `\%12\relax}%
\providecommand \@@startlink[1]{}%
\providecommand \@@endlink[0]{}%
\providecommand \url  [0]{\begingroup\@sanitize@url \@url }%
\providecommand \@url [1]{\endgroup\@href {#1}{\urlprefix }}%
\providecommand \urlprefix  [0]{URL }%
\providecommand \Eprint [0]{\href }%
\providecommand \doibase [0]{https://doi.org/}%
\providecommand \selectlanguage [0]{\@gobble}%
\providecommand \bibinfo  [0]{\@secondoftwo}%
\providecommand \bibfield  [0]{\@secondoftwo}%
\providecommand \translation [1]{[#1]}%
\providecommand \BibitemOpen [0]{}%
\providecommand \bibitemStop [0]{}%
\providecommand \bibitemNoStop [0]{.\EOS\space}%
\providecommand \EOS [0]{\spacefactor3000\relax}%
\providecommand \BibitemShut  [1]{\csname bibitem#1\endcsname}%
\let\auto@bib@innerbib\@empty
%</preamble>
\bibitem [{\citenamefont {Ceperley}\ and\ \citenamefont
  {Alder}(1986)}]{Ceperley_Science86_QMC}%
  \BibitemOpen
  \bibfield  {author} {\bibinfo {author} {\bibfnamefont {D.}~\bibnamefont
  {Ceperley}}\ and\ \bibinfo {author} {\bibfnamefont {B.}~\bibnamefont
  {Alder}},\ }\href {https://doi.org/10.1126/science.231.4738.555} {\bibfield
  {journal} {\bibinfo  {journal} {Science}\ }\textbf {\bibinfo {volume}
  {231}},\ \bibinfo {pages} {555} (\bibinfo {year} {1986})}\BibitemShut
  {NoStop}%
\bibitem [{\citenamefont {Troyer}\ and\ \citenamefont
  {Wiese}(2005)}]{Troyer_PRL05_SignProblem}%
  \BibitemOpen
  \bibfield  {author} {\bibinfo {author} {\bibfnamefont {M.}~\bibnamefont
  {Troyer}}\ and\ \bibinfo {author} {\bibfnamefont {U.-J.}\ \bibnamefont
  {Wiese}},\ }\href {https://doi.org/10.1103/PhysRevLett.94.170201} {\bibfield
  {journal} {\bibinfo  {journal} {Phys. Rev. Lett.}\ }\textbf {\bibinfo
  {volume} {94}},\ \bibinfo {pages} {170201} (\bibinfo {year}
  {2005})}\BibitemShut {NoStop}%
\bibitem [{\citenamefont {Verstraete}\ \emph {et~al.}(2008)\citenamefont
  {Verstraete}, \citenamefont {Murg},\ and\ \citenamefont
  {Cirac}}]{Verstraete_AiP08_DMRG}%
  \BibitemOpen
  \bibfield  {author} {\bibinfo {author} {\bibfnamefont {F.}~\bibnamefont
  {Verstraete}}, \bibinfo {author} {\bibfnamefont {V.}~\bibnamefont {Murg}},\
  and\ \bibinfo {author} {\bibfnamefont {J.}~\bibnamefont {Cirac}},\ }\href
  {https://doi.org/10.1080/14789940801912366} {\bibfield  {journal} {\bibinfo
  {journal} {Advances in Physics}\ }\textbf {\bibinfo {volume} {57}},\ \bibinfo
  {pages} {143} (\bibinfo {year} {2008})}\BibitemShut {NoStop}%
\bibitem [{\citenamefont {Schollwöck}(2011)}]{Schollwock_AP11_MPS}%
  \BibitemOpen
  \bibfield  {author} {\bibinfo {author} {\bibfnamefont {U.}~\bibnamefont
  {Schollwöck}},\ }\href
  {https://doi.org/https://doi.org/10.1016/j.aop.2010.09.012} {\bibfield
  {journal} {\bibinfo  {journal} {Annals of Physics}\ }\textbf {\bibinfo
  {volume} {326}},\ \bibinfo {pages} {96 } (\bibinfo {year}
  {2011})}\BibitemShut {NoStop}%
\bibitem [{\citenamefont {Becca}\ and\ \citenamefont
  {Sorella}(2017)}]{Becca_17_QMCtext}%
  \BibitemOpen
  \bibfield  {author} {\bibinfo {author} {\bibfnamefont {F.}~\bibnamefont
  {Becca}}\ and\ \bibinfo {author} {\bibfnamefont {S.}~\bibnamefont
  {Sorella}},\ }\bibinfo {title} {Variational monte carlo},\ in\ \href
  {https://doi.org/10.1017/9781316417041.006} {\emph {\bibinfo {booktitle}
  {Quantum Monte Carlo Approaches for Correlated Systems}}}\ (\bibinfo
  {publisher} {Cambridge University Press},\ \bibinfo {year} {2017})\ p.\
  \bibinfo {pages} {103–130}\BibitemShut {NoStop}%
\bibitem [{\citenamefont {Tahara}\ and\ \citenamefont
  {Imada}(2008)}]{Tahara_JPSJ08_mVMC}%
  \BibitemOpen
  \bibfield  {author} {\bibinfo {author} {\bibfnamefont {D.}~\bibnamefont
  {Tahara}}\ and\ \bibinfo {author} {\bibfnamefont {M.}~\bibnamefont {Imada}},\
  }\href {https://doi.org/10.1143/JPSJ.77.114701} {\bibfield  {journal}
  {\bibinfo  {journal} {Journal of the Physical Society of Japan}\ }\textbf
  {\bibinfo {volume} {77}},\ \bibinfo {pages} {114701} (\bibinfo {year}
  {2008})}\BibitemShut {NoStop}%
\bibitem [{\citenamefont {Carleo}\ and\ \citenamefont
  {Troyer}(2017)}]{Carleo_Science17_NQS}%
  \BibitemOpen
  \bibfield  {author} {\bibinfo {author} {\bibfnamefont {G.}~\bibnamefont
  {Carleo}}\ and\ \bibinfo {author} {\bibfnamefont {M.}~\bibnamefont
  {Troyer}},\ }\href {https://doi.org/10.1126/science.aag2302} {\bibfield
  {journal} {\bibinfo  {journal} {Science}\ }\textbf {\bibinfo {volume}
  {355}},\ \bibinfo {pages} {602} (\bibinfo {year} {2017})}\BibitemShut
  {NoStop}%
\bibitem [{\citenamefont {Dumoulin}\ and\ \citenamefont
  {Visin}(2016)}]{Dumoulin_arxiv18_IntroCNN}%
  \BibitemOpen
  \bibfield  {author} {\bibinfo {author} {\bibfnamefont {V.}~\bibnamefont
  {Dumoulin}}\ and\ \bibinfo {author} {\bibfnamefont {F.}~\bibnamefont
  {Visin}},\ }\href@noop {} {\bibinfo {title} {A guide to convolution
  arithmetic for deep learning}} (\bibinfo {year} {2016}),\ \Eprint
  {https://arxiv.org/abs/1603.07285} {arXiv:1603.07285 [stat.ML]} \BibitemShut
  {NoStop}%
\bibitem [{\citenamefont {Cohen}\ and\ \citenamefont
  {Welling}(2016)}]{Cohen_arxiv16_GCNN}%
  \BibitemOpen
  \bibfield  {author} {\bibinfo {author} {\bibfnamefont {T.~S.}\ \bibnamefont
  {Cohen}}\ and\ \bibinfo {author} {\bibfnamefont {M.}~\bibnamefont
  {Welling}},\ }\href {http://arxiv.org/abs/1602.07576} {\bibfield  {journal}
  {\bibinfo  {journal} {CoRR}\ }\textbf {\bibinfo {volume} {abs/1602.07576}}
  (\bibinfo {year} {2016})},\ \Eprint {https://arxiv.org/abs/1602.07576}
  {arXiv:1602.07576} \BibitemShut {NoStop}%
\bibitem [{\citenamefont {Nomura}(2021)}]{Nomura_JPCM2021_RBMsymm}%
  \BibitemOpen
  \bibfield  {author} {\bibinfo {author} {\bibfnamefont {Y.}~\bibnamefont
  {Nomura}},\ }\href {https://doi.org/10.1088/1361-648x/abe268} {\bibfield
  {journal} {\bibinfo  {journal} {Journal of Physics: Condensed Matter}\
  }\textbf {\bibinfo {volume} {33}},\ \bibinfo {pages} {174003} (\bibinfo
  {year} {2021})}\BibitemShut {NoStop}%
\bibitem [{\citenamefont {Choo}\ \emph {et~al.}(2019)\citenamefont {Choo},
  \citenamefont {Neupert},\ and\ \citenamefont {Carleo}}]{Choo_PRB19_J1J2CNN}%
  \BibitemOpen
  \bibfield  {author} {\bibinfo {author} {\bibfnamefont {K.}~\bibnamefont
  {Choo}}, \bibinfo {author} {\bibfnamefont {T.}~\bibnamefont {Neupert}},\ and\
  \bibinfo {author} {\bibfnamefont {G.}~\bibnamefont {Carleo}},\ }\href
  {https://doi.org/10.1103/PhysRevB.100.125124} {\bibfield  {journal} {\bibinfo
   {journal} {Phys. Rev. B}\ }\textbf {\bibinfo {volume} {100}},\ \bibinfo
  {pages} {125124} (\bibinfo {year} {2019})}\BibitemShut {NoStop}%
\bibitem [{\citenamefont {Roth}\ and\ \citenamefont
  {MacDonald}(2021)}]{Roth_arxiv21_GCNN}%
  \BibitemOpen
  \bibfield  {author} {\bibinfo {author} {\bibfnamefont {C.}~\bibnamefont
  {Roth}}\ and\ \bibinfo {author} {\bibfnamefont {A.~H.}\ \bibnamefont
  {MacDonald}},\ }\href@noop {} {\bibinfo {title} {Group convolutional neural
  networks improve quantum state accuracy}} (\bibinfo {year} {2021}),\ \Eprint
  {https://arxiv.org/abs/2104.05085} {arXiv:2104.05085 [quant-ph]} \BibitemShut
  {NoStop}%
\bibitem [{\citenamefont {Vieijra}\ \emph {et~al.}(2020)\citenamefont
  {Vieijra}, \citenamefont {Casert}, \citenamefont {Nys}, \citenamefont
  {De~Neve}, \citenamefont {Haegeman}, \citenamefont {Ryckebusch},\ and\
  \citenamefont {Verstraete}}]{Vieijra_PRL20_NQSsymm}%
  \BibitemOpen
  \bibfield  {author} {\bibinfo {author} {\bibfnamefont {T.}~\bibnamefont
  {Vieijra}}, \bibinfo {author} {\bibfnamefont {C.}~\bibnamefont {Casert}},
  \bibinfo {author} {\bibfnamefont {J.}~\bibnamefont {Nys}}, \bibinfo {author}
  {\bibfnamefont {W.}~\bibnamefont {De~Neve}}, \bibinfo {author} {\bibfnamefont
  {J.}~\bibnamefont {Haegeman}}, \bibinfo {author} {\bibfnamefont
  {J.}~\bibnamefont {Ryckebusch}},\ and\ \bibinfo {author} {\bibfnamefont
  {F.}~\bibnamefont {Verstraete}},\ }\href
  {https://doi.org/10.1103/PhysRevLett.124.097201} {\bibfield  {journal}
  {\bibinfo  {journal} {Phys. Rev. Lett.}\ }\textbf {\bibinfo {volume} {124}},\
  \bibinfo {pages} {097201} (\bibinfo {year} {2020})}\BibitemShut {NoStop}%
\bibitem [{\citenamefont {Vieijra}\ and\ \citenamefont
  {Nys}(2021)}]{Vieijra_arxiv21_NQSsymm}%
  \BibitemOpen
  \bibfield  {author} {\bibinfo {author} {\bibfnamefont {T.}~\bibnamefont
  {Vieijra}}\ and\ \bibinfo {author} {\bibfnamefont {J.}~\bibnamefont {Nys}},\
  }\bibfield  {journal} {\bibinfo  {journal} {Physical Review B}\ }\textbf
  {\bibinfo {volume} {104}},\ \href
  {https://doi.org/10.1103/physrevb.104.045123} {10.1103/physrevb.104.045123}
  (\bibinfo {year} {2021})\BibitemShut {NoStop}%
\bibitem [{\citenamefont {Sharir}\ \emph {et~al.}(2020)\citenamefont {Sharir},
  \citenamefont {Levine}, \citenamefont {Wies}, \citenamefont {Carleo},\ and\
  \citenamefont {Shashua}}]{Sharir_PRL20_QVAN}%
  \BibitemOpen
  \bibfield  {author} {\bibinfo {author} {\bibfnamefont {O.}~\bibnamefont
  {Sharir}}, \bibinfo {author} {\bibfnamefont {Y.}~\bibnamefont {Levine}},
  \bibinfo {author} {\bibfnamefont {N.}~\bibnamefont {Wies}}, \bibinfo {author}
  {\bibfnamefont {G.}~\bibnamefont {Carleo}},\ and\ \bibinfo {author}
  {\bibfnamefont {A.}~\bibnamefont {Shashua}},\ }\href
  {https://doi.org/10.1103/PhysRevLett.124.020503} {\bibfield  {journal}
  {\bibinfo  {journal} {Phys. Rev. Lett.}\ }\textbf {\bibinfo {volume} {124}},\
  \bibinfo {pages} {020503} (\bibinfo {year} {2020})}\BibitemShut {NoStop}%
\bibitem [{\citenamefont {Yang}\ \emph {et~al.}(2020)\citenamefont {Yang},
  \citenamefont {Leng}, \citenamefont {Yu}, \citenamefont {Patel},
  \citenamefont {Hu},\ and\ \citenamefont {Pu}}]{Li_PRR20_ISGO}%
  \BibitemOpen
  \bibfield  {author} {\bibinfo {author} {\bibfnamefont {L.}~\bibnamefont
  {Yang}}, \bibinfo {author} {\bibfnamefont {Z.}~\bibnamefont {Leng}}, \bibinfo
  {author} {\bibfnamefont {G.}~\bibnamefont {Yu}}, \bibinfo {author}
  {\bibfnamefont {A.}~\bibnamefont {Patel}}, \bibinfo {author} {\bibfnamefont
  {W.-J.}\ \bibnamefont {Hu}},\ and\ \bibinfo {author} {\bibfnamefont
  {H.}~\bibnamefont {Pu}},\ }\href
  {https://doi.org/10.1103/PhysRevResearch.2.012039} {\bibfield  {journal}
  {\bibinfo  {journal} {Phys. Rev. Research}\ }\textbf {\bibinfo {volume}
  {2}},\ \bibinfo {pages} {012039} (\bibinfo {year} {2020})}\BibitemShut
  {NoStop}%
\bibitem [{\citenamefont {Li}\ \emph {et~al.}(2021)\citenamefont {Li},
  \citenamefont {Liang}, \citenamefont {Chen}, \citenamefont {Xiao},
  \citenamefont {Lin}, \citenamefont {Jiang}, \citenamefont {Zhao},
  \citenamefont {Wang}, \citenamefont {He},\ and\ \citenamefont
  {An}}]{Li_arxiv21_LargeScaleCNN}%
  \BibitemOpen
  \bibfield  {author} {\bibinfo {author} {\bibfnamefont {M.}~\bibnamefont
  {Li}}, \bibinfo {author} {\bibfnamefont {X.}~\bibnamefont {Liang}}, \bibinfo
  {author} {\bibfnamefont {J.}~\bibnamefont {Chen}}, \bibinfo {author}
  {\bibfnamefont {Q.}~\bibnamefont {Xiao}}, \bibinfo {author} {\bibfnamefont
  {R.}~\bibnamefont {Lin}}, \bibinfo {author} {\bibfnamefont {Q.}~\bibnamefont
  {Jiang}}, \bibinfo {author} {\bibfnamefont {X.}~\bibnamefont {Zhao}},
  \bibinfo {author} {\bibfnamefont {F.}~\bibnamefont {Wang}}, \bibinfo {author}
  {\bibfnamefont {L.}~\bibnamefont {He}},\ and\ \bibinfo {author}
  {\bibfnamefont {H.}~\bibnamefont {An}},\ }\href@noop {} {\bibinfo {title}
  {Bridging the gap between deep learning and frustrated quantum spin system
  for extreme-scale simulations on new generation of sunway supercomputer}}
  (\bibinfo {year} {2021}),\ \Eprint {https://arxiv.org/abs/2108.13830}
  {arXiv:2108.13830 [cond-mat.str-el]} \BibitemShut {NoStop}%
\bibitem [{\citenamefont {Cai}\ and\ \citenamefont
  {Liu}(2018)}]{Cai_PRB18_CombinedNet}%
  \BibitemOpen
  \bibfield  {author} {\bibinfo {author} {\bibfnamefont {Z.}~\bibnamefont
  {Cai}}\ and\ \bibinfo {author} {\bibfnamefont {J.}~\bibnamefont {Liu}},\
  }\href {https://doi.org/10.1103/PhysRevB.97.035116} {\bibfield  {journal}
  {\bibinfo  {journal} {Phys. Rev. B}\ }\textbf {\bibinfo {volume} {97}},\
  \bibinfo {pages} {035116} (\bibinfo {year} {2018})}\BibitemShut {NoStop}%
\bibitem [{\citenamefont {Szab\'o}\ and\ \citenamefont
  {Castelnovo}(2020)}]{Szabo_PRR20_SignProblem}%
  \BibitemOpen
  \bibfield  {author} {\bibinfo {author} {\bibfnamefont {A.}~\bibnamefont
  {Szab\'o}}\ and\ \bibinfo {author} {\bibfnamefont {C.}~\bibnamefont
  {Castelnovo}},\ }\href {https://doi.org/10.1103/PhysRevResearch.2.033075}
  {\bibfield  {journal} {\bibinfo  {journal} {Phys. Rev. Research}\ }\textbf
  {\bibinfo {volume} {2}},\ \bibinfo {pages} {033075} (\bibinfo {year}
  {2020})}\BibitemShut {NoStop}%
\bibitem [{\citenamefont {Ferrari}\ \emph {et~al.}(2019)\citenamefont
  {Ferrari}, \citenamefont {Becca},\ and\ \citenamefont
  {Carrasquilla}}]{Ferrari_PRB19_GWFRBM}%
  \BibitemOpen
  \bibfield  {author} {\bibinfo {author} {\bibfnamefont {F.}~\bibnamefont
  {Ferrari}}, \bibinfo {author} {\bibfnamefont {F.}~\bibnamefont {Becca}},\
  and\ \bibinfo {author} {\bibfnamefont {J.}~\bibnamefont {Carrasquilla}},\
  }\href {https://doi.org/10.1103/PhysRevB.100.125131} {\bibfield  {journal}
  {\bibinfo  {journal} {Phys. Rev. B}\ }\textbf {\bibinfo {volume} {100}},\
  \bibinfo {pages} {125131} (\bibinfo {year} {2019})}\BibitemShut {NoStop}%
\bibitem [{\citenamefont {Nomura}\ \emph {et~al.}(2017)\citenamefont {Nomura},
  \citenamefont {Darmawan}, \citenamefont {Yamaji},\ and\ \citenamefont
  {Imada}}]{Nomura_PRB17_RBMPP}%
  \BibitemOpen
  \bibfield  {author} {\bibinfo {author} {\bibfnamefont {Y.}~\bibnamefont
  {Nomura}}, \bibinfo {author} {\bibfnamefont {A.~S.}\ \bibnamefont
  {Darmawan}}, \bibinfo {author} {\bibfnamefont {Y.}~\bibnamefont {Yamaji}},\
  and\ \bibinfo {author} {\bibfnamefont {M.}~\bibnamefont {Imada}},\ }\href
  {https://doi.org/10.1103/PhysRevB.96.205152} {\bibfield  {journal} {\bibinfo
  {journal} {Phys. Rev. B}\ }\textbf {\bibinfo {volume} {96}},\ \bibinfo
  {pages} {205152} (\bibinfo {year} {2017})}\BibitemShut {NoStop}%
\bibitem [{\citenamefont {Nomura}\ and\ \citenamefont
  {Imada}(2020)}]{Nomura_arxiv20_J1J2RBMPP}%
  \BibitemOpen
  \bibfield  {author} {\bibinfo {author} {\bibfnamefont {Y.}~\bibnamefont
  {Nomura}}\ and\ \bibinfo {author} {\bibfnamefont {M.}~\bibnamefont {Imada}},\
  }\href@noop {} {\bibinfo {title} {Dirac-type nodal spin liquid revealed by
  machine learning}} (\bibinfo {year} {2020}),\ \Eprint
  {https://arxiv.org/abs/2005.14142} {arXiv:2005.14142 [cond-mat.str-el]}
  \BibitemShut {NoStop}%
\bibitem [{\citenamefont {Liang}\ \emph {et~al.}(2021)\citenamefont {Liang},
  \citenamefont {Dong},\ and\ \citenamefont {He}}]{Liang_PRB21_CNNPEPS}%
  \BibitemOpen
  \bibfield  {author} {\bibinfo {author} {\bibfnamefont {X.}~\bibnamefont
  {Liang}}, \bibinfo {author} {\bibfnamefont {S.-J.}\ \bibnamefont {Dong}},\
  and\ \bibinfo {author} {\bibfnamefont {L.}~\bibnamefont {He}},\ }\href
  {https://doi.org/10.1103/PhysRevB.103.035138} {\bibfield  {journal} {\bibinfo
   {journal} {Phys. Rev. B}\ }\textbf {\bibinfo {volume} {103}},\ \bibinfo
  {pages} {035138} (\bibinfo {year} {2021})}\BibitemShut {NoStop}%
\bibitem [{\citenamefont {Westerhout}\ \emph {et~al.}(2020)\citenamefont
  {Westerhout}, \citenamefont {Astrakhantsev}, \citenamefont {Tikhonov},
  \citenamefont {Katsnelson},\ and\ \citenamefont
  {Bagrov}}]{Westerhout_NC20_FrustratedDifficulty}%
  \BibitemOpen
  \bibfield  {author} {\bibinfo {author} {\bibfnamefont {T.}~\bibnamefont
  {Westerhout}}, \bibinfo {author} {\bibfnamefont {N.}~\bibnamefont
  {Astrakhantsev}}, \bibinfo {author} {\bibfnamefont {K.~S.}\ \bibnamefont
  {Tikhonov}}, \bibinfo {author} {\bibfnamefont {M.~I.}\ \bibnamefont
  {Katsnelson}},\ and\ \bibinfo {author} {\bibfnamefont {A.~A.}\ \bibnamefont
  {Bagrov}},\ }\href {https://doi.org/10.1038/s41467-020-15402-w} {\bibfield
  {journal} {\bibinfo  {journal} {Nature Communications}\ }\textbf {\bibinfo
  {volume} {11}},\ \bibinfo {pages} {1593} (\bibinfo {year}
  {2020})}\BibitemShut {NoStop}%
\bibitem [{\citenamefont {Astrakhantsev}\ \emph {et~al.}(2021)\citenamefont
  {Astrakhantsev}, \citenamefont {Westerhout}, \citenamefont {Tiwari},
  \citenamefont {Choo}, \citenamefont {Chen}, \citenamefont {Fischer},
  \citenamefont {Carleo},\ and\ \citenamefont
  {Neupert}}]{Astrakhantsev_arxiv21_pyrochlore}%
  \BibitemOpen
  \bibfield  {author} {\bibinfo {author} {\bibfnamefont {N.}~\bibnamefont
  {Astrakhantsev}}, \bibinfo {author} {\bibfnamefont {T.}~\bibnamefont
  {Westerhout}}, \bibinfo {author} {\bibfnamefont {A.}~\bibnamefont {Tiwari}},
  \bibinfo {author} {\bibfnamefont {K.}~\bibnamefont {Choo}}, \bibinfo {author}
  {\bibfnamefont {A.}~\bibnamefont {Chen}}, \bibinfo {author} {\bibfnamefont
  {M.~H.}\ \bibnamefont {Fischer}}, \bibinfo {author} {\bibfnamefont
  {G.}~\bibnamefont {Carleo}},\ and\ \bibinfo {author} {\bibfnamefont
  {T.}~\bibnamefont {Neupert}},\ }\href@noop {} {\bibinfo {title}
  {Broken-symmetry ground states of the heisenberg model on the pyrochlore
  lattice}} (\bibinfo {year} {2021}),\ \Eprint
  {https://arxiv.org/abs/2101.08787} {arXiv:2101.08787 [cond-mat.str-el]}
  \BibitemShut {NoStop}%
\bibitem [{\citenamefont {Audet}\ and\ \citenamefont
  {Hare}(2017)}]{Audet_17_blackbox}%
  \BibitemOpen
  \bibfield  {author} {\bibinfo {author} {\bibfnamefont {C.}~\bibnamefont
  {Audet}}\ and\ \bibinfo {author} {\bibfnamefont {W.}~\bibnamefont {Hare}},\
  }\href@noop {} {\emph {\bibinfo {title} {Derivative-free and blackbox
  optimization}}}\ (\bibinfo  {publisher} {Springer},\ \bibinfo {year}
  {2017})\BibitemShut {NoStop}%
\bibitem [{\citenamefont {Hansen}(2016)}]{Hansen_arxiv16_CMAES}%
  \BibitemOpen
  \bibfield  {author} {\bibinfo {author} {\bibfnamefont {N.}~\bibnamefont
  {Hansen}},\ }\href@noop {} {\bibinfo {title} {The cma evolution strategy: A
  tutorial}} (\bibinfo {year} {2016}),\ \Eprint
  {https://arxiv.org/abs/1604.00772} {arXiv:1604.00772 [cs.LG]} \BibitemShut
  {NoStop}%
\bibitem [{\citenamefont {Hagym\'asi}\ \emph {et~al.}(2021)\citenamefont
  {Hagym\'asi}, \citenamefont {Sch\"afer}, \citenamefont {Moessner},\ and\
  \citenamefont {Luitz}}]{Hagymasi_PRL21_PyrochloreDMRG}%
  \BibitemOpen
  \bibfield  {author} {\bibinfo {author} {\bibfnamefont {I.}~\bibnamefont
  {Hagym\'asi}}, \bibinfo {author} {\bibfnamefont {R.}~\bibnamefont
  {Sch\"afer}}, \bibinfo {author} {\bibfnamefont {R.}~\bibnamefont
  {Moessner}},\ and\ \bibinfo {author} {\bibfnamefont {D.~J.}\ \bibnamefont
  {Luitz}},\ }\href {https://doi.org/10.1103/PhysRevLett.126.117204} {\bibfield
   {journal} {\bibinfo  {journal} {Phys. Rev. Lett.}\ }\textbf {\bibinfo
  {volume} {126}},\ \bibinfo {pages} {117204} (\bibinfo {year}
  {2021})}\BibitemShut {NoStop}%
\bibitem [{\citenamefont {Torlai}\ \emph {et~al.}(2018)\citenamefont {Torlai},
  \citenamefont {Mazzola}, \citenamefont {Carrasquilla}, \citenamefont
  {Troyer}, \citenamefont {Melko},\ and\ \citenamefont
  {Carleo}}]{Torlai_NP18_Tomography}%
  \BibitemOpen
  \bibfield  {author} {\bibinfo {author} {\bibfnamefont {G.}~\bibnamefont
  {Torlai}}, \bibinfo {author} {\bibfnamefont {G.}~\bibnamefont {Mazzola}},
  \bibinfo {author} {\bibfnamefont {J.}~\bibnamefont {Carrasquilla}}, \bibinfo
  {author} {\bibfnamefont {M.}~\bibnamefont {Troyer}}, \bibinfo {author}
  {\bibfnamefont {R.}~\bibnamefont {Melko}},\ and\ \bibinfo {author}
  {\bibfnamefont {G.}~\bibnamefont {Carleo}},\ }\href
  {https://doi.org/10.1038/s41567-018-0048-5} {\bibfield  {journal} {\bibinfo
  {journal} {Nature Physics}\ }\textbf {\bibinfo {volume} {14}},\ \bibinfo
  {pages} {447} (\bibinfo {year} {2018})}\BibitemShut {NoStop}%
\bibitem [{\citenamefont {Sorella}\ \emph {et~al.}(2007)\citenamefont
  {Sorella}, \citenamefont {Casula},\ and\ \citenamefont
  {Rocca}}]{Sorella_JCP07_SR}%
  \BibitemOpen
  \bibfield  {author} {\bibinfo {author} {\bibfnamefont {S.}~\bibnamefont
  {Sorella}}, \bibinfo {author} {\bibfnamefont {M.}~\bibnamefont {Casula}},\
  and\ \bibinfo {author} {\bibfnamefont {D.}~\bibnamefont {Rocca}},\ }\href
  {https://doi.org/10.1063/1.2746035} {\bibfield  {journal} {\bibinfo
  {journal} {The Journal of Chemical Physics}\ }\textbf {\bibinfo {volume}
  {127}},\ \bibinfo {pages} {014105} (\bibinfo {year} {2007})}\BibitemShut
  {NoStop}%
\bibitem [{\citenamefont {Hansen}\ \emph {et~al.}(2010)\citenamefont {Hansen},
  \citenamefont {Auger}, \citenamefont {Ros}, \citenamefont {Finck},\ and\
  \citenamefont {Po{\v{s}}{\'\i}k}}]{hansen_10_compare}%
  \BibitemOpen
  \bibfield  {author} {\bibinfo {author} {\bibfnamefont {N.}~\bibnamefont
  {Hansen}}, \bibinfo {author} {\bibfnamefont {A.}~\bibnamefont {Auger}},
  \bibinfo {author} {\bibfnamefont {R.}~\bibnamefont {Ros}}, \bibinfo {author}
  {\bibfnamefont {S.}~\bibnamefont {Finck}},\ and\ \bibinfo {author}
  {\bibfnamefont {P.}~\bibnamefont {Po{\v{s}}{\'\i}k}},\ }in\ \href@noop {}
  {\emph {\bibinfo {booktitle} {Proceedings of the 12th annual conference
  companion on Genetic and evolutionary computation}}}\ (\bibinfo {year}
  {2010})\ pp.\ \bibinfo {pages} {1689--1696}\BibitemShut {NoStop}%
\bibitem [{\citenamefont {Hansen}\ \emph {et~al.}(2019)\citenamefont {Hansen},
  \citenamefont {Akimoto},\ and\ \citenamefont {Baudis}}]{Hansen_19_pycma}%
  \BibitemOpen
  \bibfield  {author} {\bibinfo {author} {\bibfnamefont {N.}~\bibnamefont
  {Hansen}}, \bibinfo {author} {\bibfnamefont {Y.}~\bibnamefont {Akimoto}},\
  and\ \bibinfo {author} {\bibfnamefont {P.}~\bibnamefont {Baudis}},\ }\href
  {https://doi.org/10.5281/zenodo.2559634} {\bibinfo {title} {{CMA-ES/pycma} on
  {G}ithub}} (\bibinfo {year} {2019})\BibitemShut {NoStop}%
\bibitem [{\citenamefont {Gong}\ \emph {et~al.}(2014)\citenamefont {Gong},
  \citenamefont {Zhu}, \citenamefont {Sheng}, \citenamefont {Motrunich},\ and\
  \citenamefont {Fisher}}]{Gong_PRL14_J1J2DMRG}%
  \BibitemOpen
  \bibfield  {author} {\bibinfo {author} {\bibfnamefont {S.-S.}\ \bibnamefont
  {Gong}}, \bibinfo {author} {\bibfnamefont {W.}~\bibnamefont {Zhu}}, \bibinfo
  {author} {\bibfnamefont {D.~N.}\ \bibnamefont {Sheng}}, \bibinfo {author}
  {\bibfnamefont {O.~I.}\ \bibnamefont {Motrunich}},\ and\ \bibinfo {author}
  {\bibfnamefont {M.~P.~A.}\ \bibnamefont {Fisher}},\ }\href
  {https://doi.org/10.1103/PhysRevLett.113.027201} {\bibfield  {journal}
  {\bibinfo  {journal} {Phys. Rev. Lett.}\ }\textbf {\bibinfo {volume} {113}},\
  \bibinfo {pages} {027201} (\bibinfo {year} {2014})}\BibitemShut {NoStop}%
\bibitem [{\citenamefont {Iqbal}\ \emph {et~al.}(2019)\citenamefont {Iqbal},
  \citenamefont {M\"uller}, \citenamefont {Ghosh}, \citenamefont {Gingras},
  \citenamefont {Jeschke}, \citenamefont {Rachel}, \citenamefont {Reuther},\
  and\ \citenamefont {Thomale}}]{Iqbal_PRX19_Pyrochlore}%
  \BibitemOpen
  \bibfield  {author} {\bibinfo {author} {\bibfnamefont {Y.}~\bibnamefont
  {Iqbal}}, \bibinfo {author} {\bibfnamefont {T.}~\bibnamefont {M\"uller}},
  \bibinfo {author} {\bibfnamefont {P.}~\bibnamefont {Ghosh}}, \bibinfo
  {author} {\bibfnamefont {M.~J.~P.}\ \bibnamefont {Gingras}}, \bibinfo
  {author} {\bibfnamefont {H.~O.}\ \bibnamefont {Jeschke}}, \bibinfo {author}
  {\bibfnamefont {S.}~\bibnamefont {Rachel}}, \bibinfo {author} {\bibfnamefont
  {J.}~\bibnamefont {Reuther}},\ and\ \bibinfo {author} {\bibfnamefont
  {R.}~\bibnamefont {Thomale}},\ }\href
  {https://doi.org/10.1103/PhysRevX.9.011005} {\bibfield  {journal} {\bibinfo
  {journal} {Phys. Rev. X}\ }\textbf {\bibinfo {volume} {9}},\ \bibinfo {pages}
  {011005} (\bibinfo {year} {2019})}\BibitemShut {NoStop}%
\bibitem [{\citenamefont {Marshall}\ and\ \citenamefont
  {Peierls}(1955)}]{Marshall_PRSA55_MSR}%
  \BibitemOpen
  \bibfield  {author} {\bibinfo {author} {\bibfnamefont {W.}~\bibnamefont
  {Marshall}}\ and\ \bibinfo {author} {\bibfnamefont {R.~E.}\ \bibnamefont
  {Peierls}},\ }\href {https://doi.org/10.1098/rspa.1955.0200} {\bibfield
  {journal} {\bibinfo  {journal} {Proceedings of the Royal Society of London.
  Series A. Mathematical and Physical Sciences}\ }\textbf {\bibinfo {volume}
  {232}},\ \bibinfo {pages} {48} (\bibinfo {year} {1955})}\BibitemShut
  {NoStop}%
\bibitem [{\citenamefont {Liang}\ \emph {et~al.}(2018)\citenamefont {Liang},
  \citenamefont {Liu}, \citenamefont {Lin}, \citenamefont {Guo}, \citenamefont
  {Zhang},\ and\ \citenamefont {He}}]{Liang_PRB18_CNNJ1J2}%
  \BibitemOpen
  \bibfield  {author} {\bibinfo {author} {\bibfnamefont {X.}~\bibnamefont
  {Liang}}, \bibinfo {author} {\bibfnamefont {W.-Y.}\ \bibnamefont {Liu}},
  \bibinfo {author} {\bibfnamefont {P.-Z.}\ \bibnamefont {Lin}}, \bibinfo
  {author} {\bibfnamefont {G.-C.}\ \bibnamefont {Guo}}, \bibinfo {author}
  {\bibfnamefont {Y.-S.}\ \bibnamefont {Zhang}},\ and\ \bibinfo {author}
  {\bibfnamefont {L.}~\bibnamefont {He}},\ }\href
  {https://doi.org/10.1103/PhysRevB.98.104426} {\bibfield  {journal} {\bibinfo
  {journal} {Phys. Rev. B}\ }\textbf {\bibinfo {volume} {98}},\ \bibinfo
  {pages} {104426} (\bibinfo {year} {2018})}\BibitemShut {NoStop}%
\bibitem [{\citenamefont {Glorot}\ \emph {et~al.}(2011)\citenamefont {Glorot},
  \citenamefont {Bordes},\ and\ \citenamefont
  {Bengio}}]{Glorot_AISTATS11_ReLU}%
  \BibitemOpen
  \bibfield  {author} {\bibinfo {author} {\bibfnamefont {X.}~\bibnamefont
  {Glorot}}, \bibinfo {author} {\bibfnamefont {A.}~\bibnamefont {Bordes}},\
  and\ \bibinfo {author} {\bibfnamefont {Y.}~\bibnamefont {Bengio}},\ }in\
  \href@noop {} {\emph {\bibinfo {booktitle} {Proceedings of the fourteenth
  international conference on artificial intelligence and statistics}}}\
  (\bibinfo {year} {2011})\ pp.\ \bibinfo {pages} {315--323}\BibitemShut
  {NoStop}%
\bibitem [{\citenamefont {Choo}\ \emph {et~al.}(2018)\citenamefont {Choo},
  \citenamefont {Carleo}, \citenamefont {Regnault},\ and\ \citenamefont
  {Neupert}}]{Choo_PRL18_SymNet}%
  \BibitemOpen
  \bibfield  {author} {\bibinfo {author} {\bibfnamefont {K.}~\bibnamefont
  {Choo}}, \bibinfo {author} {\bibfnamefont {G.}~\bibnamefont {Carleo}},
  \bibinfo {author} {\bibfnamefont {N.}~\bibnamefont {Regnault}},\ and\
  \bibinfo {author} {\bibfnamefont {T.}~\bibnamefont {Neupert}},\ }\href
  {https://doi.org/10.1103/PhysRevLett.121.167204} {\bibfield  {journal}
  {\bibinfo  {journal} {Phys. Rev. Lett.}\ }\textbf {\bibinfo {volume} {121}},\
  \bibinfo {pages} {167204} (\bibinfo {year} {2018})}\BibitemShut {NoStop}%
\bibitem [{\citenamefont {M\"unger}\ and\ \citenamefont
  {Novotny}(1991)}]{Munger_PRB91_ReweightingMC}%
  \BibitemOpen
  \bibfield  {author} {\bibinfo {author} {\bibfnamefont {E.~P.}\ \bibnamefont
  {M\"unger}}\ and\ \bibinfo {author} {\bibfnamefont {M.~A.}\ \bibnamefont
  {Novotny}},\ }\href {https://doi.org/10.1103/PhysRevB.43.5773} {\bibfield
  {journal} {\bibinfo  {journal} {Phys. Rev. B}\ }\textbf {\bibinfo {volume}
  {43}},\ \bibinfo {pages} {5773} (\bibinfo {year} {1991})}\BibitemShut
  {NoStop}%
\bibitem [{\citenamefont {Nakamura}\ \emph {et~al.}(1992)\citenamefont
  {Nakamura}, \citenamefont {Hatano},\ and\ \citenamefont
  {Nishimori}}]{Nakamura_JPSJ92_ReweightingQMC}%
  \BibitemOpen
  \bibfield  {author} {\bibinfo {author} {\bibfnamefont {T.}~\bibnamefont
  {Nakamura}}, \bibinfo {author} {\bibfnamefont {N.}~\bibnamefont {Hatano}},\
  and\ \bibinfo {author} {\bibfnamefont {H.}~\bibnamefont {Nishimori}},\ }\href
  {https://doi.org/10.1143/JPSJ.61.3494} {\bibfield  {journal} {\bibinfo
  {journal} {Journal of the Physical Society of Japan}\ }\textbf {\bibinfo
  {volume} {61}},\ \bibinfo {pages} {3494} (\bibinfo {year}
  {1992})}\BibitemShut {NoStop}%
\bibitem [{\citenamefont {Nesterov}(2003)}]{Nesterov_03}%
  \BibitemOpen
  \bibfield  {author} {\bibinfo {author} {\bibfnamefont {Y.}~\bibnamefont
  {Nesterov}},\ }\href@noop {} {\emph {\bibinfo {title} {Introductory lectures
  on convex optimization: A basic course}}},\ Vol.~\bibinfo {volume} {87}\
  (\bibinfo  {publisher} {Springer Science \& Business Media},\ \bibinfo {year}
  {2003})\BibitemShut {NoStop}%
\bibitem [{\citenamefont {Kingma}\ and\ \citenamefont
  {Ba}(2017)}]{Kingma_17_Adam}%
  \BibitemOpen
  \bibfield  {author} {\bibinfo {author} {\bibfnamefont {D.~P.}\ \bibnamefont
  {Kingma}}\ and\ \bibinfo {author} {\bibfnamefont {J.}~\bibnamefont {Ba}},\
  }\href@noop {} {\bibinfo {title} {Adam: A method for stochastic
  optimization}} (\bibinfo {year} {2017}),\ \Eprint
  {https://arxiv.org/abs/1412.6980} {arXiv:1412.6980 [cs.LG]} \BibitemShut
  {NoStop}%
\end{thebibliography}%
\clearpage

\appendix

\begin{figure}[t]
    \centering
    \includegraphics[width=0.48\textwidth]{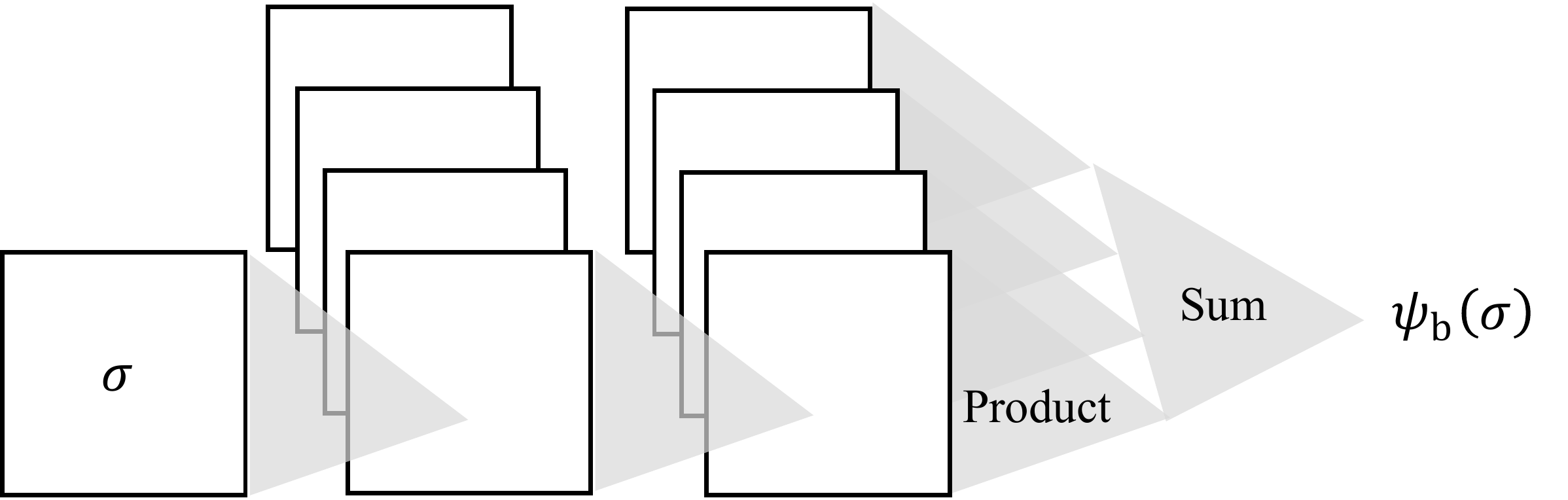}
    \caption{Base network. The first convolutional layer has bias and ReLU activation. The second convolutional layer has configuration product activation without bias.}
    \label{fig:base_network}
\end{figure}

\section{Base network architecture} \label{base network}
The full base network is a two-layer real-valued CNN with a \texttt{ReLU} activation \cite{Glorot_AISTATS11_ReLU} in the first layer and a \textit{configuration product} activation in the second layer, as shown in Fig.\,\ref{fig:base_network}. In the configuration product activation function, every channel of the convolutional output is transformed into a number given by the product of all elements in the channel, and the sum over these numbers gives the network output $\psi_{\mathrm{b}}$. The name ``configuration product'' comes from the pair product states~\cite{Tahara_JPSJ08_mVMC} which sum over the products of pair entanglement factors to give the wave function. The full base network expresses $\psi_{\mathrm{b}}$ instead of $\log\psi_{\mathrm{b}}$, which permits the expression of sign structures without complex output values. The similar architectures are also used in some other NQS studies~\cite{Liang_PRB21_CNNPEPS, Li_arxiv21_LargeScaleCNN} and greatly improve the accuracy. In our numerical practice, the training of the base network is also faster compared with the networks expressing $\log\psi_{\mathrm{b}}$. In the case of the pyrochlore lattice, the channel number of both layers are set to 20 in the base network so that the total number of parameters in the base network and the sign network is similar to Ref.\,\cite{Astrakhantsev_arxiv21_pyrochlore}.

In this work, we would need an amplitude base network with strictly positive outputs. To this end, the \texttt{ShiftedReLU} activation, defined as
\begin{equation} \label{eq:ShiftedReLU}
    \mathrm{ShiftedReLU}(x) = \mathrm{ReLU}(x) + 1 = \left\{
    \begin{array}{cc}
        x+1 & (x>0) \\
        1 & (x<0)
    \end{array} \right. ,
\end{equation}
is added before the configuration product activation function. 

\section{Symmetry} \label{symmetries}

Symmetry plays an important role in the practice of NQS \cite{Nomura_JPCM2021_RBMsymm, Choo_PRL18_SymNet}. In the numerical experiments we apply suitable symmetry projectors to increase the accuracy of variational states. With the $q = 0$ translational symmetry already enforced by the CNN structure, we only discuss the remaining point group symmetries. Assuming the system permits a symmetry group of order $\nu$ represented by operators ${T_i}$ with characters ${\omega_i}$, the symmetrized wave function is then defined as
\begin{equation} \label{eq:wave_function_symm}
\begin{aligned}
    \psi^{\mathrm{symm}}(\sigma) &= \frac{1}{\nu} \sum_i \omega_i^{-1} \psi(\hat T_i \sigma) \\ 
    &= \frac{1}{\nu} \sum_i \omega_i^{-1} \psi_{\mathrm{b}}(\hat T_i \sigma) \times \psi_{\mathrm{h,sign}}(\hat T_i \sigma),
\end{aligned}
\end{equation}
with the original wave function $\psi(\sigma)$ given by Eq.\,\eqref{eq:wave_function}. 

\begin{figure}[t]
    \centering
    \includegraphics[width=0.48\textwidth]{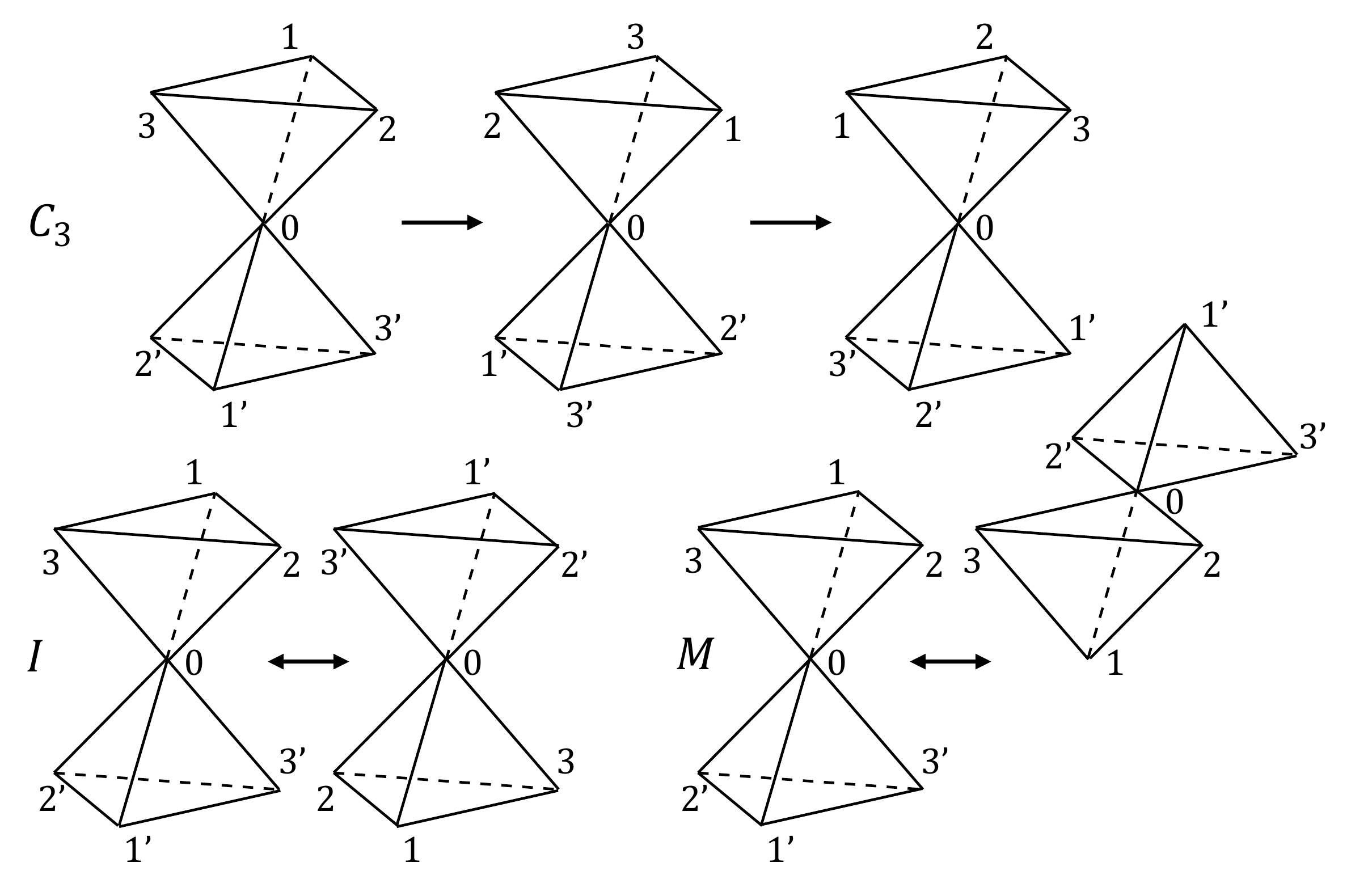}
    \caption{Symmetries of the pyrochlore lattice. $C_3$ is the three-fold rotation symmetry around the ``easy axis''. $I$ is the inversion symmetry with respect to the origin. $M$ is the mirror symmetry with the mirror plane passing the bond 2--3 and the middle point of 0 and 1.}
    \label{fig:pyrochlore_symm}
\end{figure}

At the pyrochlore lattice, we apply the $C_3$ rotation symmetry around the ``easy axis'', $I$ inverse symmetry with respect to the origin and $M$ mirror symmetry shown in Fig.\,\ref{fig:pyrochlore_symm}. Those generators cover the full point symmetry group of the pyrochlore cluster. The ground state belongs to the trivial representation of the resulting symmetry group.

\section{Reweighting Monte Carlo} \label{reweighting}

Similarly to QMC, reweighting Monte Carlo has been applied to the study of statistical systems \cite{Munger_PRB91_ReweightingMC,Nakamura_JPSJ92_ReweightingQMC}. In this section we use the evaluation of energy as an example to illustrate the reweighting method.

To begin with, we define the two norms of the non-normalized quantum state $\ket{\Psi}$
\begin{equation}
    ||\Psi|| = \sqrt{\braket{\Psi|\Psi}} = \sqrt{\sum_\sigma|\psi_\sigma|^2},
\end{equation}
\begin{equation}
    |\Psi| = \sum_\sigma|\psi_\sigma|.
\end{equation}

In the variational Monte Carlo approach, the energy is given by
\begin{equation} \label{eq:variational_energy2}
\begin{aligned}
    E &= \sum_{\sigma,\sigma'}\frac{\psi_\sigma^*}{||\Psi||}\frac{\psi_{\sigma'}}{||\Psi||}H_{\sigma \sigma'} \\
    &= \sum_\sigma \frac{|\psi_\sigma|^2}{||\Psi||^2} \sum_{\sigma'} \frac{\psi_{\sigma'}}{\psi_\sigma}H_{\sigma \sigma'}\\
    &= \left\langle \sum_{\sigma'}\frac{\psi_\sigma'}{\psi_\sigma}H_{\sigma\sigma'} \right\rangle_2,
\end{aligned}
\end{equation}
where $H_{\sigma \sigma'} = \braket{\sigma|H|\sigma'}$, and $\left\langle \ldots \right\rangle_2$ denotes sampling with the weight $|\psi_{\sigma}|^2$. 

The reweighting Monte Carlo estimates the energy as
\begin{equation}
\begin{aligned}
    E &= \frac{|\Psi|}{||\Psi||^2} \sum_\sigma \frac{|\psi_\sigma|}{|\Psi|} \frac{|\psi_\sigma|}{\psi_\sigma} \sum_{\sigma'}\psi_{\sigma'} H_{\sigma\sigma'} \\
    &= \frac{|\Psi|}{||\Psi||^2} \left\langle \frac{|\psi_\sigma|}{\psi_\sigma} \sum_{\sigma'}\psi_{\sigma'} H_{\sigma\sigma'} \right\rangle_1,
\end{aligned}
\end{equation}
where $\left\langle \ldots \right\rangle_1$ denotes sampling with the weight $|\psi_\sigma|$. The factor $|\Psi|/||\Psi||^2$ is given by
\begin{equation}
    \frac{|\Psi|}{||\Psi||^2} = \frac{|\Psi|}{\sum_\sigma |\psi_\sigma|^2} 
    = \frac{1}{\sum_\sigma \frac{|\psi_\sigma|}{|\Psi|}|\psi_\sigma|} 
    = \frac{1}{\left\langle |\psi_\sigma| \right\rangle_1}.
\end{equation}
Finally, the energy in the reweighting VMC reads
\begin{equation} \label{eq:reweighting_energy}
    E = \frac{1}{\left\langle |\psi_\sigma| \right\rangle_1} \left\langle \frac{|\psi_\sigma|}{\psi_\sigma} \sum_{\sigma'} \psi_{\sigma'} H_{\sigma \sigma'} \right\rangle_1.
\end{equation}
More generally,
\begin{equation}
    \left\langle \ldots \right\rangle_2 
    = \frac{\left\langle |\psi_\sigma| \ldots \right\rangle_1}{\left\langle |\psi_\sigma| \right\rangle_1},
\end{equation}
which can help to write the SR optimization in a reweighting form.

When $|\psi_{\sigma'}|$ is much greater than $|\psi_\sigma|$, in VMC without reweighting the $\psi_{\sigma'}/\psi_\sigma$ ratio in Eq.\,\eqref{eq:variational_energy2} leads to large variance in the energy estimation, and the evolution strategy updates will focus on a few largest terms in the sampling, leading to unstable optimization. The reweighting method is introduced mainly for ameliorating this problem.

The reweighting Monte Carlo also makes it easier to update wave function entries with small amplitudes since the reweighting sampling probability is proportional to $|\psi|$ instead of $|\psi|^2$. Consequently, the reweighting Monte Carlo is adopted across all numerical experiments in this paper except calculating cumulative distribution functions to improve the accuracy of variational wave function.

\section{Covariance matrix adaptation evolution strategy (CMA-ES) details} \label{CMA-ES}

\textit{Hyperparameters.--\,}
The CMA-ES contains many hyperparameters. In numerical experiments, we set the ES sample number $N=100$, initial weights $\bm W = 0$ and the initial standard deviation of sampling $\sigma=\pi/4$. The default values in the \texttt{pycma} package~\cite{Hansen_19_pycma} are adopted for remaining hyperparameters. 

The update weights $w_n$ are also chosen as the default values in the \texttt{pycma} package. In a descending order, they are given by
\begin{equation}
    w_n' = \log \frac{N+1}{2} - \log n,
\end{equation}
\begin{equation}
    w_n = \left\{
    \begin{array}{rc}
        w_n' / \sum |w_n'|^+,           & w_n' \geq 0 \\
        \alpha^- w_n' / \sum |w_n'|^-,  & w_n' < 0
    \end{array}
    \right.,
\end{equation}
where $\sum |w_n'|^+$ and $\sum |w_n'|^-$ are the sum over all positive and negative $w_n'$ values, respectively. Here $\alpha^-$ is a multiplier introduced in Ref.\,\cite{Hansen_arxiv16_CMAES}, helping the information from previous iterations accelerate the optimization in a manner similar to the momentum term in gradient-based methods~\cite{Nesterov_03, Kingma_17_Adam}.

\textit{Estimation of energy in ES.--\,} An essential step in ES is the estimation of the variational energy $E_n$ for different sample weights $X_n$ through Eq.\,\eqref{eq:reweighting_energy}. To simplify the estimation, we use the same reweighting Monte Carlo samples for all sample networks because the Monte Carlo sampling probability is not influenced by the sign of the wave function if symmetries in Eq.\,\eqref{eq:wave_function_symm} are not applied. The base network outputs $\psi_{\mathrm{b}}(\sigma)$ and $\psi_{\mathrm{b}}(\sigma')$ are prepared for calculating the variational energy. These prepared values can also be used in the following SR step. After applying symmetries, the sign network is no longer trained. This simplification does not cause a significant loss of accuracy according to our numerical experiments.

Given $N$ sample networks in every iteration, an efficient calculation requires the parallel forward pass of all these networks. In practice we only define one sample network with its output channel number $N$ times of the original sign network so that it produces $N$ sign values corresponding to different sample weights in a single forward pass.

\textit{Comparison with simple ES.--\,} In the simple ES, the correlation of weights is not considered in sample generation and the weights with lowest variational energy are directly chosen as the new weights in the sign network. In our numerical experiments, the CMA-ES, as a popular variant of ES method, produces sign structures with lower variational energies compared with simple ES.

\begin{figure}[t]
    \centering
    \subfigure{ \label{fig:supplementary_comparison_energy}
        \includegraphics[width=0.23\textwidth]{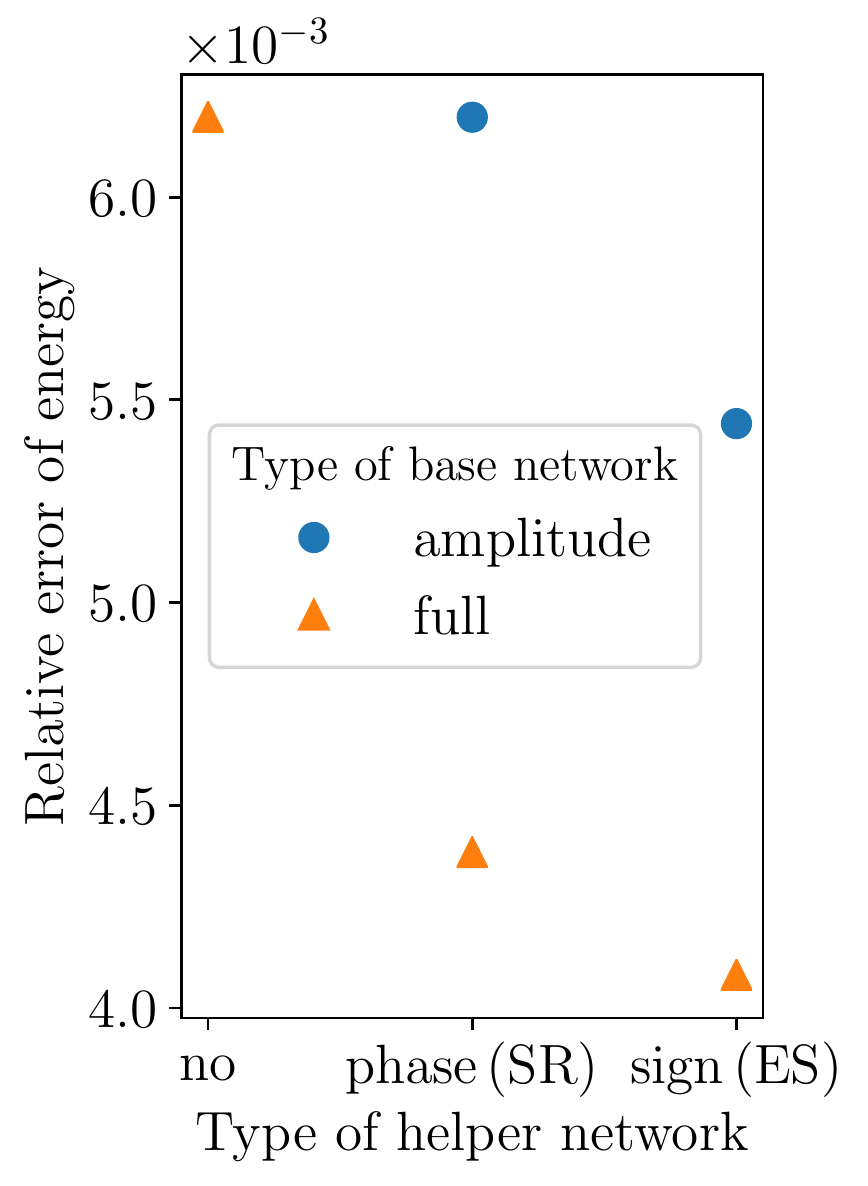}
    }
    \hspace{-0.6cm}
    \subfigure{ \label{fig:supplementary_comparison_overlap}
        \includegraphics[width=0.24\textwidth]{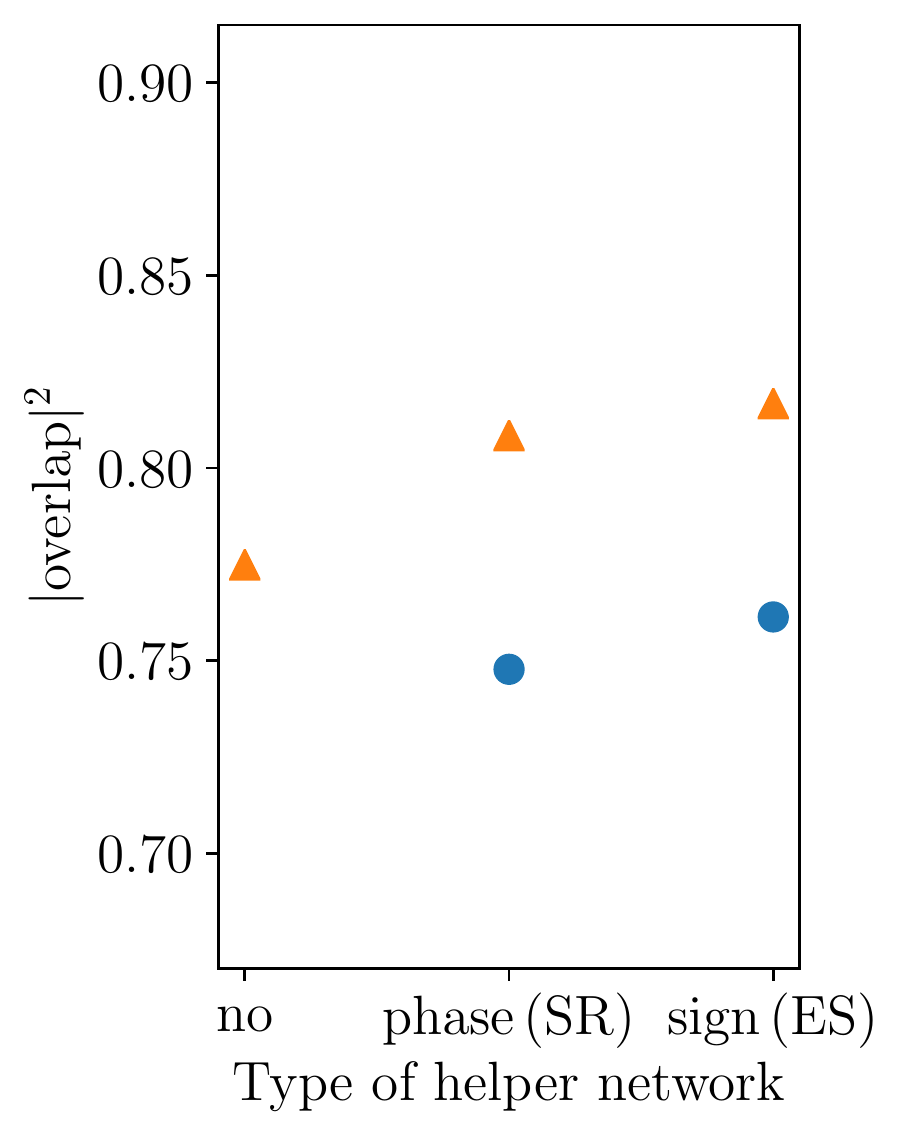}
    }
    \caption{Comparison between different base and helper networks on the $4\times2^3$ pyrochlore Heisenberg model.}
    \label{fig:pyrochlore_supplementary}
\end{figure}

\begin{figure}[t]
    \centering
    \includegraphics[width=0.45\textwidth]{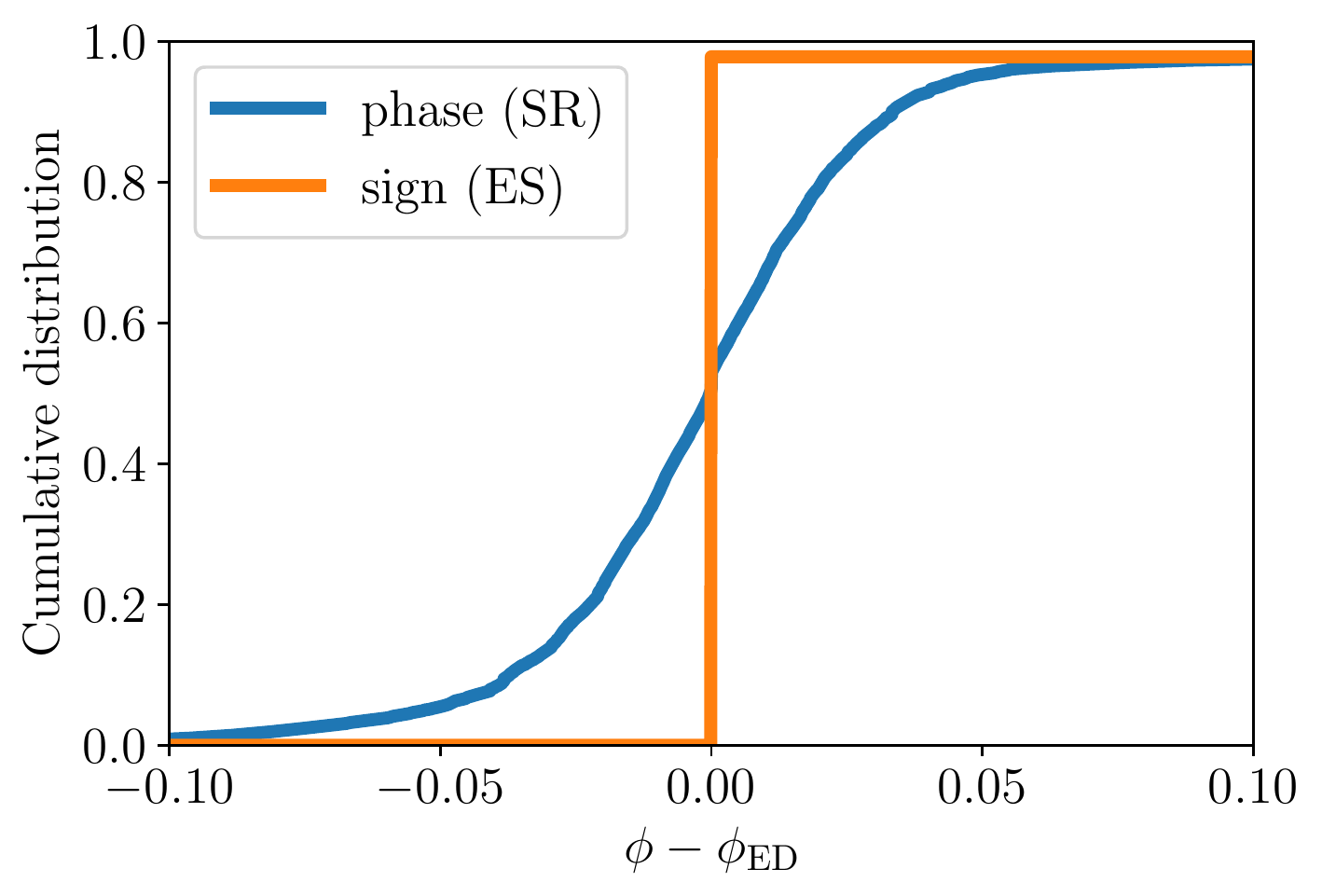}
    \caption{Cumulative distribution function (CDF) of phase relative to ED on $4\times2^3$ pyrochlore lattice. The full network is combined with the phase or sign helper network. The phase is adjusted to lie within the range $(-\pi/2, 3\pi/2]$.}
    \label{fig:pyrochlore_CDF}
\end{figure}

\section{Supplementary data for pyrochlore lattice} \label{supplementary_pyrochlore}

In Fig.\,\ref{fig:pyrochlore_supplementary}, we show the comparison of the energy error and the wave function overlap between different base networks and helper networks. The sign helper network outperforms the phase helper network, and the full base network also outperforms the amplitude base network.

Similar to Fig.\,\ref{fig:square_CDF}, in Fig.\,\ref{fig:pyrochlore_CDF} we depict the CDF of the variational phase obtained within the combination of the full base network and different helper networks relative to the phase $\phi_{\mathrm{ED}}$ expected within the ED. The CDF shows that the sign network trained by ES learns a better distribution of phases due to elimination of a  redundant complex phase.

\section{Training details}

\textit{Training procedure.--\,} The training procedure is designed to ensure that the sign network expresses the primary sign structure such as MSR and the base network expresses the remaining sign structure. The whole training procedure contains the following steps:
\begin{enumerate}[itemsep=0pt, topsep=0pt]
    \item No base network is used and all amplitudes are set to 1. The sign network is trained by the ES until the variational energy is converged.
    \item The base network is put into the combined network structure and trained by the SR until the variational energy stabilizes.
    \item The sign network is again trained by the ES.
    \item Both networks are trained simultaneously for several thousands of iterations to convergence.
    \item Point symmetry group defined in Appendix~\ref{symmetries} are applied and the base network is trained by the SR.
\end{enumerate}

\textit{Adjustment of base network weights.--\,} 
Two possible problems may occur in the base network. Firstly, the output of the base network may overflow since the network expresses $\psi_{\mathrm{b}}$ directly instead of $\log\psi_{\mathrm{b}}$. Secondly, the scale of parameters may be significantly different in different layers, which leads to vanishing or exploding gradients.

Fortunately, one can adjust the weights of the full base network to alleviate the two problems. The architecture of the full base network in Fig.\,\ref{fig:base_network} contains two redundant degrees of freedom. One is free to multiply a number to all variational parameters in the first or second layer. This transformation essentially multiplies an overall constant to the wave function without changing the physical state.

Assuming one multiplies a constant $A$ to the weights in the first layer ${\bm W}^{(1)}$ and another constant $B$ to the weights in the second layer ${\bm W}^{(2)}$, then the transformation can be written in short as
\begin{equation} \label{eq:transform_weight}
    {\bm W}^{(1)} \rightarrow A {\bm W}^{(1)}, \quad {\bm W}^{(2)} \rightarrow B {\bm W}^{(2)}.
\end{equation}
This transformation also modifies the neurons in the first layer ${\bm v}^{(1)}$, the neurons in the second layer ${\bm v}^{(2)}$ and the network output $\psi$ by
\begin{equation} \label{eq:transform_neuron}
    {\bm v}^{(1)} \rightarrow A {\bm v}^{(1)}, \quad {\bm v}^{(2)} \rightarrow AB {\bm v}^{(2)}, \quad \psi \rightarrow (AB)^n \psi,
\end{equation}
where $n$ is the number of elements in each channel. After each SR step, we hope the maximal wave function entry $M$ in Monte Carlo samples can be normalized to 1, so
\begin{equation} \label{eq:normalize_M}
    (AB)^n M = 1.
\end{equation}

On the other hand, the standard deviation of weights also undergoes similar transformation
\begin{equation}
    \sigma^{(1)} \rightarrow A\sigma^{(1)}, \quad
    \sigma^{(2)} \rightarrow B\sigma^{(2)},
\end{equation}
where $(1)$ and $(2)$ still represent different layers. In our numerical experiments, the network has best performance when $\sigma^{(1)}$ and $\sigma^{(2)}$ are kept equal, so we define their ratio
\begin{equation}
    R = \frac{\sigma^{(1)}}{\sigma^{(2)}} 
    \rightarrow \frac{A}{B} R,
\end{equation}
which is measured in every iteration and modified to 1, giving
\begin{equation} \label{eq:normalize_R}
    \frac{A}{B} R = 1
\end{equation}

Combining Eq.\,\eqref{eq:normalize_M} and Eq.\,\eqref{eq:normalize_R}, one obtains
\begin{equation}
    A = \sqrt{\frac{1}{R M^{1/n}}}, \quad B = \sqrt{\frac{R}{M^{1/n}}}.
\end{equation}
We apply the transformation Eq.\,\eqref{eq:transform_weight} with the above $A$ and $B$ values to ensure the stability and efficiency of the training process.

\end{document}